%% file: ms.tex
\documentstyle[12pt,aaspp4,psfig]{article}

\lefthead{T. Heckman et al.}
\righthead{Absorption-Lines in Galactic Winds}
 
\slugcomment{.}
 
\begin{document}
 
\title{Absorption-Line Probes of Gas and Dust in Galactic Superwinds}
 
\author{Timothy M. Heckman$^1$}
\affil{Department of Physics and Astronomy, Johns Hopkins
University, Homewood Campus, 3400 North Charles Street, Baltimore, MD 21218}
\author{Matthew D. Lehnert$^1$}
\affil{Max-Plank-Institut f\"ur extraterrestrische Physik, Postfach 1603,
D-85740 Garching, Germany}
\author{David K. Strickland}
\affil{Department of Physics and Astronomy, Johns Hopkins
University, Homewood Campus, 3400 North Charles Street, Baltimore, MD 21218}
\and
\author{Lee Armus$^1$}
\affil{SIRTF Science Center, 310-6, Caltech, Pasadena, CA 91125}
 
\parindent=0em
\vspace{4cm}
 
1. Visiting astronomers, Kitt Peak National Observatory and Cerro Tololo
Interamerican Observatory, NOAO,
operated by AURA, Inc. under cooperative agreement with the
National Science Foundation.

\newpage
 
\parindent=2em
 
\begin{abstract}
We have 
obtained moderate resolution ($R$ = a few thousand) spectra of
the $NaI\lambda\lambda$5890,5896 ($NaD$) absorption-line in
a sample of 32 far-IR-bright starburst galaxies.
In 18 cases,
the $NaD$ line in the nucleus is produced primarily by interstellar gas,
while cool stars
contribute significantly in the others.
In 12 of the 18 ``interstellar-dominated'' cases
the $NaD$ line is blueshifted by over 100
km s$^{-1}$ relative to the galaxy systemic velocity (the
``outflow sources''), while no case shows a net
{\it redshift} of more than 100 km s$^{-1}$. 
The absorption-line profiles in these outflow sources
span the range from near the galaxy systemic velocity
to a maximum blueshift
of $\sim$ 400 to 600 km s$^{-1}$. The outflow
sources are galaxies systematically viewed more nearly face-on than
the others. We therefore argue that the absorbing
material consists of ambient interstellar material that has been 
entrained and accelerated along the minor axis of the galaxy by 
a hot starburst-driven superwind. The
$NaD$ lines are optically-thick, but indirect arguments imply
total Hydrogen column densities of $N_H \sim$ few
$\times 10^{21}$ cm$^{-2}$. This implies that
the superwind is expelling matter
at a rate comparable to the star-formation rate. 
This outflowing material
is evidently very dusty: we
find a strong correlation between the depth of the $NaD$ profile
and the line-of-sight reddening.
Typical implied values
are $E(B-V)$ = 0.3 to 1 over regions several-to-ten kpc in size.
We briefly consider some of the potential implications of these observations.
The estimated terminal velocities of superwinds inferred from the
present data and extant X-ray data are typically 400 to 800 km s$^{-1}$,
are independent of the galaxy rotation speed, and are comparable to 
(substantially exceed) the escape
velocities for $L_*$ (dwarf) galaxies. The resulting selective loss of metals
from shallower potential wells can
establish the mass-metallicity relation in spheroids,
produce the observed metallicity in the intra-cluster medium,
and enrich a general IGM to of-order 10$^{-1}$ solar metallicity.
If the outflowing dust grains can survive their journey into the IGM, 
their effect on observations
of cosmologically-distant objects would be significant. 
\end{abstract}
 
 
\newpage
 
 
\section{Introduction}

By now, it is well-established that galactic-scale outflows of gas (sometimes
called `superwinds') are a ubiquitous phenomenon in the most actively star-
forming galaxies in the local universe (Heckman, Lehnert,
\& Armus 1993; Dahlem 1997;
Bland-Hawthorn 1995). They are  powered by the energy deposited in the
interstellar medium by massive stars via supernovae and stellar winds.
Over the history of the universe, outflows like these may
have polluted the intergalactic medium with metals (e.g. Giroux \& Shull
1997) and dust (Alton, Davis, \& Bianchi 1999; Aguirre 1999a,b), heated and
polluted the intracluster medium
(e.g. Gibson, Loewenstein, \& 
Mushotzky 1997; Ponman, Cannon, \& Navarro 1999),
and may have established the mass-metallicity relation and radial metallicity
gradients in galactic spheroids  (e.g. Carollo \& Danziger 1994).
However,
the astrophysical relevance of superwinds can not be
reliably assessed without first understanding their  physical, dynamical, and
chemical properties. To date, most of the pertinent information has come
from observations of  the X-ray emission produced by the hot gas (e.g.
Dahlem, Weaver, \& Heckman 1998) or the optical line-emission produced
by the warm gas (e.g. Lehnert \& Heckman 1996a).

In the present paper, we take a complementary approach, and discuss an
extensive body of new data that probes  the outflowing gas via its interstellar
absorption-lines. This technique has some important advantages. First, since
the gas is seen in absorption against the background starlight, there is no
possible ambiguity as to the sign (inwards or outwards) of any radial flow
that is detected. Second, the strength of the absorption will be related  to the
column density of the gas. In contrast, the X-ray or optical surface-brightness
of the emitting gas is proportional to the emission-measure. Thus, the
absorption-lines more fully probe the whole range of gas densities in the
outflow, rather than being strongly weighted in favor of the densest material
(which may contain relatively little mass).  Finally, provided that suitably-
bright background sources can be found, interstellar absorption-lines can be
used to study outflows in high-redshift galaxies where the associated X-ray
or optical {\it emission} may be undetectably faint. This promise has already
been realized in the case of the 'Lyman Dropout' galaxies, where the
kinematic signature of outflows is clear in their rest-frame UV spectra
(Franx et al 1997; Pettini et al 1998, 1999).

A few pioneering studies have already detected interstellar absorption-lines
from superwinds in local starburst galaxies. Phillips (1993) discussed 
spatially-resolved optical spectroscopy of the $NaI$ ``D'' doublet in NGC 1808,
showing that an outflow of gas at velocities of  up to 700 km s$^{-1}$ could
be traced over a region several kpc in size, coincident with a region of extra-
planar dust plumes. Several recent papers (Lequeux et al. 1995; Heckman \&
Leitherer 1997; Sahu \& Blades 1997; Kunth et al. 1998; Gonzalez-Delgado
et al 1998a) have detected blueshifted interstellar absorption-lines in $HST$
and $HUT$
UV spectra of a handful of starburst galaxies, implying outflows of metal-
bearing gas at velocities of 10$^2$ to 10$^3$ km s$^{-1}$.

Most of the strong resonance lines of cosmically-abundant ions are found in
the UV (e.g. Morton 1991; Savage \& Sembach 1996),
and so must be studied with $HST$ or $FUSE$
in local starbursts. In the present program, we have instead exploited the
relatively greater sensitivity and availability of ground-based telescopes at
visible wavelengths to study a large sample of starbursts/superwinds using
the $NaI$ doublet at $\lambda$$\lambda$5890,5896 \AA. In a few cases,
we have also observed the $KI$ $\lambda$$\lambda$7665,7699 \AA\ doublet, since
it probes gas with nearly the same ionization state as $NaI$, but is more
likely to be optically thin (the $K$ abundance is down from $Na$ by a factor
of 15, while the two doublets have similar oscillator strengths). The
ionization potentials of $NaI$ and $KI$ are only 5.1 eV and 4.3 eV
respectively, so these species should primarily probe the $HI$ and $H_2$
ISM phases. Observations in vacuum-UV will be required to study the hotter
and more highly ionized gas in absorption.

\section{The Data}

\subsection{Sample Selection}

Table 1 lists the salient properties of  the 32 objects in our sample. These
objects have been drawn  from the larger samples of infrared-selected
galaxies studied by Armus, Heckman, \& Miley (1989 - hereafter AHM) and
by Lehnert \& Heckman (1995 - hereafter LH95). The specific selection
criteria used in these in two programs are described in detail in these
references. Briefly, AHM selected  on the basis of far-IR flux and very warm
far-IR color-temperatures. LH95 selected on the basis of far-IR flux,
moderately warm far-IR color-temperatures, and high galaxy inclination (
disk galaxies seen within $\sim$30$^{\circ}$ of edge-on). The AHM and LH95
samples overlap in galaxy properties, but the former preferentially selects
more powerful and more distant objects.

AHM measured the equivalent widths of the $NaD$ lines in their sample
using low-resolution spectra. The galaxies from AHM were selected for the
present program on the basis of the brightness of their nucleus at $\sim$
5900\AA\ and the equivalent width of $NaD$. The galaxies from LH95 had no
prior measures of the $NaD$ line, and were selected based primarily on their
proximity and availability at the time of the observations. 

Of the 32 objects, only 3 are classified as {\it bona fide} AGN of the basis
of their optical spectra: the type 2 Seyferts NGC7582 and Mrk273 and the
highly peculiar AGN IRAS11119+3257. No published classification exists
for IRAS10502-1843. The remaining objects are
optically classified as $HII nuclei$ or $LINER's$, and are presumed to be
primarily powered by dusty starbursts (Lutz, Veilleux, \& Genzel 1999). 
Even in the Seyfert galaxies NGC7582 and Mrk273, optical spectra show the
presence of a young stellar population in the nucleus (Schmitt, Storchi-
Bergmann, \& Cid-Fernadez 1999; Gonalzez-Delgado, Heckman, \&
Leitherer 2000). Thus, with the possible exception of IRAS11119+3257
and IRAS10502-1843, the
objects in our sample all contain powerful starbursts that can drive superwinds.

\subsection{Observations}

The observations were undertaken during the period from 1988 through 1994
using three different facilities: the 4-meter Blanco Telescope with the
Cassegrain Spectrograph at $CTIO$, the 4-meter Mayall Telescope with the
RC Spectrograph at $KPNO$, and the 2.5-meter Dupont Telescope with the
Modular Spectrograph at the Las Campanas Observatory. Various
spectrograph configurations were used at each observatory, the details of
which are listed in Table 2. The spectral resolution used to study the 
$NaD$ lines ranged from 55 to
170 km s$^{-1}$. While low by the standards of interstellar absorption-line
studies, the resolution was good enough to cleanly resolve the $NaD$ lines
in most cases (deconvolved line widths of 100 to 600 km s$^{-1}$ - see
below).

\subsection{Data Reduction \& Analysis}

The spectra were all processed using the standard {\it LONGSLIT} package
in $IRAF$.  All the data were bias-subtracted using the overscan region of
the chip and then flat-fielded using observations of either a quartz-lamp-
illuminated screen inside the dome or of a quartz lamp inside the
spectrograph. The spectra were then rectified  using observations of bright
stars to determine and remove the distortion perpendicular to the dispersion
direction and observations of a $HeNeAr$ arc lamp to determine the two-
dimensional dispersion solution. The zero-points in the wavelength scale
were verified  by measuring the wavelengths of strong night-sky emission-
lines. Corrections to the heliocentric reference-frame were computed for the
spectra. 
The spectra were then sky-subtracted by
interactively fitting a low-order  polynomial along the spatial direction,
column-by-column. For a few of the galaxies we obtained relatively
low-dispersion spectra
that will be used to measure the reddening (Balmer decrement and continuum
color). 
These
data were flux-calibrated using observations of spectrophotometric
standard stars, and otherwise were reduced in the same way as the other data.

The spectra were analyzed  using the interactive {\it SPLOT} spectral fitting
package in $IRAF$. In all cases, a one-dimensional `nuclear' spectrum was
extracted, covering a region with a size set by the slit width and summed over
3 to 5 pixels
in the spatial direction (typically 2 by 3 arcsec). The
corresponding linear size of the projected aperture is generally a few hundred
to a few thousand parsecs in these galaxies (median diameter 700 pc). 
This is a reasonable match to the typical sizes of powerful starbursts
like these (e.g. Meurer et al 1997; Lehnert \& Heckman 1996b).
Prior to further analysis, each 1-D spectrum was normalized to unit intensity by
fitting it with, and then dividing it by, a low-order polynomial. These nuclear
spectra are shown in Figure 1. Similar one-dimensional spectra for off-
nuclear regions were extracted over the spatial region with adequate signal-
to-noise in the continuum for each galaxy. The primary focus of the present
paper is on the nuclear spectra, but we will describe the results obtained in
the off-nuclear bins when these are particularly illuminating or interesting.

Given the relatively low resolution of our spectra, the likelihood that the
observed $NaD$ line profile contains many unresolved and/or blended kinematic
sub-components, and the saturated nature of the $NaD$ lines, we have
chosen to parameterize the lines as simply as possible. Thus, for each
extracted spectrum, we have fit the $NaD$ doublet with a single pair of
Gaussians, constrained to have the same line width and a wavelength
separation appropriate to the redshifted doublet. In a few objects, the
adjacent 
$HeI\lambda$5876 nebular emission-line was strong and broad enough to
slightly contaminate the blue half of the $NaD\lambda$5890 profile. For
these cases, we first fit and subtracted the $HeI$ 
emission line.
Only the parameters of the stronger member of $KI$ doublet at 7665 \AA\
were measured. The weaker member at 7699 \AA\ was detected, with a
strength consistent with the doublet being optically-thin (i.e. an equivalent
width ratio of $\sim$2:1).

We have not attempted a rigorous determination of the measurement uncertainties
associated with these data. The relatively high signal-to-noise in
the nuclear spectra (typically better than 30:1 per pixel) means
that the uncertainties in the measured quantities will be
dominated in most cases by systematic effects due to the contamination of the
$NaD$ line by weaker stellar photospheric features
(whose ubiquity likewise
makes it difficult to determine the true continuum level to use
in the line-fitting) and by the mismatch in profile shape between the
actual data and the single Gaussian
component used to fit each member of the doublet (see section 3.2 below). 
The most straightforward way to estimate the measurement
uncertainties is to compare
the measurements for the 11 galaxies in the sample for which we have more
than one independent spectrum (taken at a different position angle).
We have done so, and the results are reported in the Notes to Table 3.

\section{Results}

\subsection{The Stellar {\it vs.} Interstellar Contribution}

Before using the $NaD$ line to diagnose conditions in the starburst galaxies,
it is imperative to establish that the line is primarily interstellar
in origin in
these galaxies. The $NaD$ line is strong in the spectra of cool stars, reaching
a peak strength in the range from K3 through M0 (see Jacoby, Hunter, \&
Christian 1984). These stellar types can make a significant contribution to the
optical spectrum of a starburst galaxy. First, the oldest  underlying population
in the galaxy bulge will have a dominant contribution from K -type giants
(indeed the $NaD$ line is one of the strongest stellar absorption-lines in
optical spectra of early-type galaxies and bulges - e.g. Heckman 1980; Bica
et al 1991). Second, for starbursts with ages greater than about 10 Myr, cool
supergiants make a significant contribution to the optical and near-IR light
(Bruzual \& Charlot 1993; Leitherer et al 1999).

We have therefore tried to estimate empirically what fraction of the measured
equivalent width of the $NaD$ doublet is contributed  by late-type stars in
our sample galaxies. To do so, we have considered other absorption-lines that
are conspicuous in the spectra of late-type stars and galactic nuclei, but which
arise from highly-excited states and are therefore of purely stellar origin
(i.e. they are not resonance lines like $NaD$). The best-studied example is the
$MgI$ b-band at 5174\AA. The strength of this line is well-correlated with
the strength of the $NaD$ line in spectra of the nuclei of normal galaxies (
Bica et al 1991; Heckman 1980) and in stars (Jacoby, Hunter, \& Christian
1984). We have used the latter two data sets to determine a best-fit  to the
correlation: $W_{NaD}$ $\sim$ 0.75 $W_{Mg-b}$.  The measured strength
of $Mg-b$ in our galaxies (from AHM, Veilleux et al 1995, or
our own unpublished spectra) was then used  to predict the equivalent width
of the {\it stellar} contribution to the observed $NaD$ line. We have also
compared our data to spectra of K giant stars obtained during the same
observing runs listed in Table 2. Rather than measuring the strengths of a few
particular stellar features, we have used the entire ensemble of features in the
range between about 5750 and 6450 \AA\ to estimate by-eye the fractional
stellar contribution to the $NaD$ line. The agreement between these two
methods is generally satisfactory (the predicted $NaD$ stellar equivalent
widths agree on-average to $\sim$ 0.1 dex). Heckman \& Lehnert (2000)
have measured the fraction of the red continuum contributed by cool
stars for the seven nuclei in the present sample
having the highest quality detection of the {\it interstellar}
component of the $NaD$ line. They find that this fraction is 20 to 30\%,
consistent with the rough estimates reported here.

As listed in Table 3, the estimated stellar contribution to the observed $NaD$
line in our sample galaxies ranges from negligible ($<$10\%) to substantial
($>$70\%), with hints of a bimodal distribution. Thus, rather than attempting
a very uncertain direct correction for the effects of the stellar contribution,
we have taken the simpler approach of dividing our sample into two bins: the
strong-stellar-contamination objects (`SSC') in which stars produce $\geq$40\% 
of the measured $NaD$ equivalent-width, and the interstellar-dominated
objects (`ISD') in which the stellar contribution is $\leq$ 30\%. In the
discussion to follow, we will see that the $NaD$ lines in the two sub-samples
have significantly different properties, which can be readily understood as
reflecting the relative importance of the stellar and interstellar components.

\subsection{Kinematics}

The most robust indicator of an outflow is the presence of interstellar
absorption-lines that are significantly blueshifted with respect the the
systemic velocity of the galaxy ($v_{sys}$). Thus, we have first compiled the
best available measures of $v_{sys}$ for our galaxies
The velocities of the
{\it nuclear} emission-lines are potentially affected by radial gas flows and
are not always reliable indicators of $v_{sys}$ (Mirabel \& Sanders 1988;
Lehnert \& Heckman 1996a). We have therefore determined $v_{sys}$ from 
(in order of preference) spatially-resolved galactic rotation curves, global
mm-wave $CO$ line profiles, nuclear stellar velocities, global
$HI\lambda$21cm {\it emission-line} profiles, and optical nuclear emission-
line velocities (only used for 4 objects). See Table 1 for details and
the estimated uncertainties.

The results are shown in Figure 2. For the $ISD$ subsample there is strong
trend for the centroid of the $NaD$ feature to be blueshifted with respect to
$v_{sys}$. Specifically, 11 of the 18 $ISD$ nuclei have $NaD$ blueshifts
$\Delta$$v$ greater than 100 km s$^{-1}$ (hereafter the `outflow sources'). 
In addition, while the nuclear $NaD$ absorption-line in NGC1808 lies close
to $v_{sys}$, the galaxy exhibits strongly blueshifted absorption over a
several-kpc-scale region along its minor axis (Phillips 1993). We therefore
include it as a 12th member of the outflow sample. The net blueshifts in the
$ISD$ nuclei are in the range $\Delta$$v$ $\sim$ 100 to 300 km s$^{-1}$,
(with the exception of IRAS11119+3257).
\footnote{IRAS11119+3257 has perhaps the most peculiar optical spectrum of
any ultra-luminous system. It shows very broad (1500 km s$^{-1}$)
Balmer, [OIII]$\lambda$$\lambda$4959,5007, FeII, HeI, and
[OI]$\lambda$6300 emission-lines.
It appears to be a member of the ``I~Zw~1'' class of quasars (e.g. Phillips
1976), or possibly related to Mrk 231.
It is very compact (barely resolved) in optical images
(Armus, Heckman, \& Miley 1987). The $NaD$ absorption profile is complex,
with a strong narrow system that is blueshifted by 934 km s$^{-1}$,
and a weaker system blueshifted by 1410 km s$^{-1}$. See
Table 3 and Figure 1.} 
In contrast to these large blueshifts, no net {\it redshifts}
greater
than 100 km s$^{-1}$ are observed in the $ISD$ sample. Moreover,  none of
the 14 members of  the $SSC$ sample show a net $NaD$ blueshift or redshift
that is greater than 70 km s$^{-1}$. This is consistent with expectations that
the velocity of the nuclear stellar $NaD$ component will be very close to
$v_{sys}$.

The $NaD$ linewidths in the $ISD$ and $SSC$ subsamples are also
significantly different (Figure 3). The lines are relatively narrow 
in the $SSC$ subsample ($W
\sim$ 100 to 300 km s$^{-1}$, with a median of 180 km s$^{-1}$),
and much broader
in the $ISD$ nuclei ($W$ $\sim$ 150 to 600 km s$^{-1}$, with a median
of 425 km s$^{-1}$). The lines are
especially broad  (typically 400 to 600 km s$^{-1}$) in the outflow sources.
As shown in Figure 4, the net blueshift in these sources is typically about
half the line width ($\Delta$$v \sim$ 1/2 $W$). The peculiar AGN
IRAS11119+3257, with $W << \Delta$$v$, is the
notable exception. Thus, in a typical outflow, the redmost absorption occurs
close to $v_{sys}$. This is strongly suggestive of  a flow in which
matter is injected at roughly zero velocity and then accelerates outward. The
approximate implied terminal velocity of the flow is then $v_{term} \sim$
$\Delta$$v + 0.5W$, which ranges from 220 to 1450 km s$^{-1}$ in our
sample (Table 4). This picture is quite different
from the standard one of a simple expanding `superbubble' in which
the absorption is due to a thin layer of cooled post-shock gas, and
for which $W << \Delta$$v$ would be expected (e.g. Weaver et al. 1977).

It is instructive to compare the observed velocities in the absorbing material
to the velocities expected from purely gravitational forces in the starburst
galaxy. This is shown in Figure 5, where we plot $W$ {\it vs.}
the galaxy rotation
speed ($v_{rot}$ - 
see Table 1 for details). This figure has several interesting
implications. First, neither the sample as-a-whole nor any of the above
subsamples show any correlation between the velocity dispersion in the
absorbing material and the galaxy rotation speed. This suggests that gravity
does not play a dominant role in determining the dynamics of the
absorbing gas.

Figure 5 also shows that the $NaD$ lines are surprisingly narrow in the
$SSC$ sources compared with expectations for either stars or gas in the
bulge of  the starburst `host' galaxy. The lines are exceptionally 
narrow if they are stellar in origin, since in this case the observed
line broadening ($W_{obs}$) will be produced by both the intrinsic stellar line
broadening ($W_*$) and that produced by galactic dynamics ($W_{gal}$):
$W_{gal} = \sqrt{W_{obs}^2 - W_*^2}$. The observed equivalent
widths of the $Na\lambda$5890 line are in the range 2.45$\pm$0.4 \AA\
in the $SSC$ nuclei. If the absorption were purely stellar,
the {\it minimum} required
values for $W_*$ would be 125$\pm$20 km s$^{-1}$ (corresponding to
completely black stellar lines). The typical implied values for
$W_{gal}$ in the $SSC$ sample would then be
60 to 200 km s$^{-1}$, with a median value of 130 km s$^{-1}$.

To emphasize how narrow the lines are in the $SSC$ sample, we show in
Figure 5 the empirical relation (Whittle 1992; Franx 1993) between the galaxy
rotation speed and the bulge velocity dispersion as a function of
Hubble type for a sample of normal disk
galaxies. The values for $W_{obs}$ in the $SSC$ objects are on-average
$\sim$ 0.2 dex below this relation for normal galaxies of the same rotation
speed and
Hubble type (typically Sa to Sc), while the implied values for $W_{gal}$
would be even more discrepant (see above).
Put another way, based on the Hubble
types and the galaxy absolute magnitudes ($M_B$
$\sim$ -19 to -21) for the $SSC$ subsample, the Faber-
Jackson relation for normal galactic bulges would predict typical values
of $W_{gal}$ $\sim$ 200 to 300 km s$^{-1}$ (e.g. Nelson \& Whittle
1996), while the observed widths are typically only 140 to 200 km s$^{-1}$,
even without a correction for the line broadening due to $W_*$.

The nebular {\it emission} lines are also narrow in the $SSC$ nuclei, as
has been shown to be more generally true for starbursts  by Weedman (1983). 
In this case, Lehnert \& Heckman (1996b) showed that the narrowness of the
nuclear emission-lines could be understood because the ionized gas was
rotationally supported and did not fairly sample the galaxy rotation curve (it
lies within the region of the galaxy with solid-body rotation). If this
explanation applies to the $NaD$ lines in the $SSC$ nuclei, it implies that
a significant fraction of
the stellar contribution comes from a dynamically-cold (disk/starburst)
component rather than from the bulge.

Finally, Figure 5 shows that the $NaD$ linewidths are relatively large in the
outflow sources ($W$ $\sim$ 1 to 3 $v_{rot}$). As we have argued above,
the kinematic properties of the $NaD$ profiles suggest that gas is `loaded'
into the outflow at $v \sim v_{sys}$ and is then accelerated up to some
terminal velocity that corresponds to the most-blueshifted part of the $NaD$
line profile. We plot $v_{term}$ {\it vs.} $v_{rot}$ in Figure 6, from which
it is clear that $v_{term}$ is significantly larger than $v_{rot}$,
but is uncorrelated with it. This suggests that the outflows {\it may} be able
to selectively escape the shallower galactic potential wells, as we will 
discuss in section 4.2 below.

Neither the $SSC$ nor the $ISD$ subsamples show a significant correlation
between the widths of the $NaD$ absorption-line and the $H\alpha$
emission-line. In particular, the outflow sources with very broad (400 to 600
km s$^{-1}$) $NaD$ absorption-lines have $H\alpha$ emission-line widths
ranging from 145 km
s$^{-1}$ (NGC7552) to 1500 km s$^{-1}$ (IRAS11119+3257). This
presumably means that the dynamics of the more tenuous outflowing absorbing 
gas is
largely decoupled from that of the dense (high emission-measure) gas that
provides most of the nuclear line-emission.

As described above, we have fit the profile of the $NaD$ doublet with a single
pair of Gaussians constrained to have the same widths and a fixed
separation. Inspection of Figure 1 clearly shows that the observed profiles
of many of the $ISD$ sample are more complex than this. The $ISD$ profiles 
generally have a larger kurtosis than a Gaussian (i.e. narrower core and
broader wings) and are sometimes asymmetric with a weak blueward
wing on the $\lambda$5890 profile (e.g. NGC 1808, IRAS 10565+2448,
IRAS 11119+3257, NGC 6240), and/or definite substructure 
(e.g. NGC 1614, NGC 3256, IRAS 11119+3257). Observations at higher
spectral resolution should prove instructive.

\subsection{The Roles of Luminosity and Geometry}

Of the 32 galaxies in our sample, 14 show relatively weak {\it interstellar}
$NaD$
absorption-lines (the SSC sample, in which the {\it stellar} contribution to the
line is strong), 6 have predominantly interstellar $NaD$ lines lying
close to the
systemic velocity of the galaxy, and 12 have interstellar lines that
are blueshifted by more than 100 km s$^{-1}$ relative to $v_{sys}$.
These 12 outflow sources differ systematically
from the other objects in two striking
respects: they are 
more luminous starbursts and they are preferentially located in galaxies
seen relatively face-on.

Specifically, 64\% (9/14) of the 
galaxies with $L_{IR} >$ 10$^{11}$ L$_{\odot}$ show outflows, compared to only
28\% (5/18) of the less luminous galaxies. The mean values for $logL_{IR}$ are
11.44$\pm$0.18 and 10.86$\pm$0.13 for the outflow and other sources
respectively, a difference that is significant at the 2.6 $\sigma$ level.
The relationship to galaxy inclination is stronger:
69\% (11/16) of the galaxies
with a ratio of semi-major to semi-minor axes $a/b \leq$ 2.0 show outflows,
while this is true for only 6\% (1/16) of the flatter (more highly inclined)
galaxies. The mean values for $log(a/b)$ are
0.20$\pm$0.03 and 0.42$\pm$0.03 for the outflow and other sources
respectively, a difference that is significant at the 4.6 $\sigma$ level.

It is likely that the primary correlation is between an observed outflow
and low galaxy inclination (small $a/b$). The weaker apparent correlation
with $L_{IR}$ is probably induced by the 
loose anti-correlation in our sample between $L_{IR}$ and $a/b$. This anti-
correlation 
reflects our selection of galaxies from both the LH95 `edge-on' galaxy 
sample (large $a/b$ and moderate $L_{IR}$) and the AHM `FIR-warm' sample
(broad range in $a/b$ and large $L_{IR}$).

Taken at face value, the correlation with galaxy inclination implies
that there is a high probability ($\sim$70\%) that an observer located
within $\sim$60$^{\circ}$ of the rotation axis of a starburst galaxy will
see outflowing gas in absorption. This geometrical constraint is 
consistent with the observed loosely-collimated outflows seen in emission
along the minor axes of edge-on starburst galaxies (e.g. Dahlem, Weaver,
\& Heckman 1998).

\subsection{Column Densities and Optical Depths}

The $NaD$ line is clearly optically-thick in these galaxies. The ratio of the
equivalent widths of the $\lambda$5890 and $\lambda$5896 members of the
doublet ($R$) can be used to estimate the optical depth (e.g. Spitzer 1968). 
The distribution of $R$ is markedly different in the $SSC$ and $ISD$
subsamples. In the former, there is a very narrow observed range ($R$
$\sim$ 1.1 to 1.3).  This is consistent with a strong stellar contribution to
the
$NaD$ line, since $R$ $\sim$ 1.0  to 1.3 (indicative of large optical depths)
is characteristic of cool stars. The
range is much broader for the $ISD$ sample, from $R$ = 1.1 to 1.7. This
range corresponds to central optical depths in the $\lambda$5896 line of
$\tau \sim$ 20 to 0.5.

At first sight, it might appear odd that the $NaD$ line is optically-thick,
yet is
not  black at line center. This can be seen for the $ISD$ sample in Figure 7,
where we have plotted $R$ {\it vs.} the normalized residual intensity at the
center of the
$\lambda$5890 feature: $I_{5890}  = F_{5890}/ F_{cont}$ (with the respective
fluxes measured at line center and in the adjacent continuum). There is a
broad range in $I_{5890}$ from 0.14 (nearly black) to 0.7. More tellingly,
there is no correlation between $R$ and $I_{5890}$ . This implies that the
absorbing gas does not fully cover the background continuum light, and that
$I_{5890}$ is determined more by this covering factor ($C_f$) than by the
optical depth. A covering factor less than unity is natural in these galaxies.
First,
the continuum light may arise in part from stars in the galaxy that are
located in
front of most of the absorbing gas (i.e. this is not the idealized case of a
purely foreground absorbing screen: the gas and stars are likely to be mixed).
Secondly, the gas is likely to be quite clumpy and inhomogeneous (e.g.
Calzetti 1997; Gordon, Calzetti, \& Witt 1997).

In the limit of large optical depth,
$C_f = (1 - I_{5890})$, but for low or moderate optical
depth $C_f > (1 - I_{5890}$). For the typical optical depths in this sample,
we can approximate $C_f$ by (1 - $I_{5890}$).
This can be demonstrated quantitatively for those members of
the $ISD$ sample in which the $NaD$ lines are well-resolved, narrow enough
so that the two doublet members are cleanly separated from one-another
($W <$ 300 km s$^{-1}$), and that have high signal-to-noise spectra. These 
constraints leave us with only three objects: NGC1808, NGC2146, and M82.
Following Hamann et al. (1997) and Barlow \& Sargent (1997), we have:

\begin{equation}
C_f = (I_{5896}^2 - 2I_{5896} + 1)/(I_{5890} - 2I_{5896} + 1)
\end{equation}

where $I_{5896}$ is the normalized intensity at the center of the 
$\lambda$5896
line. The measured values of $C_f$ are 0.83, 0.84 and 0.84 for
NGC1808, NGC2146, and M82 respectively, while the corresponding values for
$(1 - I_{5890})$ are 0.83, 0.82, and 0.82.

In this circumstance - in which optically-thick gas only partially covers the
continuum source - the measured equivalent width of the $NaD$ doublet
($EQ$) will be insensitive to the $NaI$ column density, and will instead be
primarily determined by the product of $C_f$ and the line-of-sight
velocity dispersion in the gas.
We plot the separate dependences of $EQ$ on
$I_{5890}$ and $W$ in Figures 8 and 9 respectively for the $ISD$ sample.
It is clear from these two figures that $EQ$ is determined largely by the
covering factor (Figure 8), since there is no correlation between $W$ and
$EQ$ (Figure 9).

Given that the $NaD$ doublet is moderately optically-thick in these
galaxies, it is not straightforward to estimate a $NaI$ column density
($N_{NaI}$). We have taken three approaches, and emphasize that these are
designed to give us only a rough (order-of-magnitude) estimate. Our
techniques can potentially underestimate $N_{NaI}$, because they are
insensitive to any $NaI$ sub-component that is highly optically-thick, yet
kinematically quiescent.

The first is the classical doublet ratio method (e.g. Spitzer 1968), which
relates $R$ directly to the optical depth at line center, and thereby allows the
column density to be deduced from the equivalent width. In the spirit of this
analysis, we will not attempt to measure columns for all the individual cases,
but will instead estimate a characteristic value based on the typical observed
parameters. The median value observed in the $ISD$ sample is $R \sim$ 1.2,
implying that the corresponding median optical depth at the center of the
$NaD\lambda$5896 line is $\tau_{5896}$ $\sim$ 4 
(see Table 2.1 in Spitzer 1968).
The median
observed value $EQ \sim$ 6 \AA\ for the doublet, equation 2-41 and Table
2.1 in Spitzer (1968), and the oscillator strength from Morton (1991),
together imply $N_{NaI}$ $\sim$ 10$^{14}$ cm$^{-2}$. Note that this assumes
$C_f$ = 1, and should be increased by $C_f^{-1}$, or a typical
factor of $\sim$ 1.6. 

A variant of the doublet-ratio technique can be applied to the three cases
discussed above
in which the two members of the $NaD$ doublet are cleanly separated
and well-resolved (NGC1808, NGC2146, and M82). Again, following 
Hamann et al (1997) we have:

\begin{equation}
\tau_{5896} = ln[C_f/(I_{5896} + C_f -1)]
\end{equation}

The resulting values for $\tau_{5896}$ are 2.3, 2.1, and 1.9 for
NGC1808, NGC2146, and M82 respectively. These are smaller than
the values implied by $R$ by a factor of $\sim$ 2 in these cases. 
The implied values for
$N_{NaI}$ are 1.0 $\times$ 10$^{14}$, 6 $\times$ 10$^{13}$,
and 6 $\times 10^{13}$ cm$^{-2}$ after
correction by $C_f^{-1}$.

We have also measured the equivalent width of the $KI\lambda$7665 line
in three of the nuclei (NGC1614, NGC1808, and NGC3256). Since $KI$ and
$NaI$ have very similar ionization potentials, and since $K$ and $Na$ show
similar grain depletion patterns (Savage \& Sembach 1996), the expected
ratio of the $NaI$ and $KI$ column densities should be 15 for gas with a
solar Na/K ratio. 
The measured values for the $NaD$ doublet ratio imply optical depths at
the center of the $NaD\lambda$5890 line of 8, 16, and 1.6 for NGC1614,
NGC1808, and NGC3256 respectively. The implied optical depths for the
$KI\lambda$7665 line would then be 0.5, 1.1, and 0.1 respectively. Using the
oscillator strength tabulated by Morton (1991),  the measured equivalent
widths of the line imply that $N_{KI}$ = 3 $\times$ 10$^{12}$, 4 $\times$
10$^{12}$, and 1.3 $\times$ 10$^{12}$ cm$^{-2}$ respectively. Assuming
$N_{NaI}$ = 15 $N_{KI}$, the corresponding $NaI$ columns are 4.5 $\times$
10$^{13}$, 6 $\times$ 10$^{13}$, and 2 $\times$ 10$^{13}$
cm$^{-2}$. These values are about a factor of two or three smaller than
would have been deduced for these three cases using the $NaD$ doublet
ratio alone. The value for NGC1808 is in good agreement with that derived
from Equation 2.
Under the circumstances, we regard the agreement between the three methods
as satisfactory, and
conclude that the typical value in the $ISD$ sample is $logN_{NaI}$ =
13.5 to 14. A final indirect indication that these $NaI$
column densities are roughly correct comes from the detections
of the ``Diffuse Interstellar Bands'' in the seven highest-quality
spectra of the $ISD$ sample (Heckman \& Lehnert 2000). The observed
strengths of these features agree with the strengths seen in Galactic
sight-lines with $logN_{NaI} \sim$ 13.5 to 14. 

What is the {\it total} gas column density associated with the outflow? To
calculate this directly from the (already uncertain) $NaI$ column requires
knowing the metallicity of the gas, the fractional depletion of $Na$ onto
grains (typically a factor of $\sim 10$ in diffuse clouds in the Milky Way)
and the potentially substantial ionization correction to account for ionized
$Na$. Assuming solar $Na$ abundances and a factor of ten correction for
depletion onto grains (e.g Savage \& Sembach 1996),
$N_{NaI}$ = 10$^{14}$ cm$^{-2}$ implies a typical
value for $N_H$ of 5 $\times$ 10$^{20}$ ($N_{Na}/N_{NaI}$)  cm$^{-2}$.
We can also take an empirical approach suggested by the correlation between
$N_{NaI}$ and the total gas column towards stars in our own Galaxy. Using
the data in Herbig (1993), values for $N_{NaI}$ in the range we estimate
($logN_{NaI} \sim$ 13.5 to 14) correspond to sight-lines with
$N_H$ $\sim$ 1.5 to 4 $\times$ 10$^{21}$ cm$^{-2}$. Interestingly,
this is just the range of values for $N_H$ deduced from the amount
of reddening along the line-of-sight to these nuclei based on either the
Balmer decrement or the colors of the optical continuum, assuming
a normal Galactic extinction-curve and dust-to gas ratio (
section 3.5, and see also AHM; Veilleux et
al 1995).

These estimates suggest that the ionization correction factor is significant but
not huge (i.e. $N_{Na}/N_{NaI} \sim$ 3 to 10). Since its ionization potential
is only 5.1 eV, the presence of relatively significant amounts of $NaI$ implies that it is
associated with gas having a significant dust optical depth in the near-UV:
for a Galactic extinction curve and dust-to-gas ratio, a Hydrogen column
density of $N_H = 8 \times 10^{20}$ cm$^{-2}$ is required to produce
$\tau_{dust} = 1$ at 5.1 eV ($\lambda \sim$ 2420 \AA).

It is instructive to compare the total column densities we infer for the
outflows of a few $\times$ 10$^{21}$ cm$^{-2}$ to the column densities
in the other components of the ISM in these galaxies. Column densities to the
nucleus for the hot X-ray-emitting gas are estimated to be of-order
10$^{21}$ cm$^{-2}$ in superwinds  (e.g. Suchkov et al 1994; Heckman et
al 1999; Strickland 1998). In the nuclei themselves, the dominant ISM component
is molecular, and the inferred columns range from $\sim$ 10$^{23}$ to
10$^{25}$ cm$^{-2}$ (e.g. Sanders \& Mirabel 1996).

The $HI\lambda$21cm line is observed in absorption against the bright
nonthermal radio sources in starburst nuclei (e.g. Koribalski 1996; Heckman
et al 1983; Mirabel \& Sanders 1988). The implied column densities are
typically a few $\times$ 10$^{21}$ to 10$^{22} (T_{spin}/100K)$ cm$^{-
2}$. The absorption is centered close to $v_{sys}$ and spans a velocity
range similar to that of the molecular gas. This strongly suggests that this gas
is a trace atomic component in the starbursting molecular disk or ring. 
The kinematics of the gas responsible for the $\lambda$21cm absorption are
therefore quite distinct from the gas that produces the blueshifted $NaD$
absorption. This has several plausible explanations. First, the outflowing
$HI$ is probably too hot ($T > 10^3$ K) to produce strong absorption at
$\lambda$21cm. Second, the background radio continuum source against which
the gas that produces the $\lambda$ 21cm absorption is observed will almost
certainly be invisible in the optical: it lies behind a total column density
(overwhelmingly $H_2$) of $\sim$ 10$^{23}$ to 10$^{25}$ cm$^{-2}$,
corresponding to $A_V$ = 60 to 6000! Clearly, such material will not
contribute to the observed $NaD$ absorption-lines.

\subsection{Dust Associated with the Absorbing Gas}

We have argued in section 3.4 that the $NaD$ lines are
are optically thick, and that $EQ$ is set primarily by 
the covering fraction for the absorbing gas ($C_f$) 
rather than by the line width ($W$ - see Figures 8 and 9).
This inference helps explain the otherwise puzzling correlations found by
AHM and Veilleux et al (1995) between $EQ$ and the reddening inferred
from either the Balmer decrement or the color of the optical continuum
in the nuclear spectra of large samples of starbursts.
For $\tau_{NaD}$ $>>$ 1 and $C_f$ = 1, $EQ$ would be set
by $W$, and so no correlation with the reddening would be expected. If
instead $EQ$ is principally determined by the fraction of the starburst that is
covered by gas containing $NaI$ (and dust grains), then this correlation is
more reasonable. 

Veilleux et al (1995) have also shown (via the Balmer decrement)
that the region of significant reddening extends far beyond the
nucleus in many far-IR-bright galaxies.
Thus, to gain further insight into the relationship between the
$NaD$ absorption
and dust-reddening, we have mapped out the spatial variation in the
depth of the $NaD$ line ($I_{5890}$) and the reddening
in the six galaxies in our $ISD$ sub-sample for which we have the relevant
data on the reddening (M82, NGC3256, NGC6240, Mrk273, IRAS03514+1546,
and IRAS10565+2448). The size of the region mapped was set by the
detectability of the $NaD$ line, and ranges from
3 to 9 kpc (except for M 82, where the mapped region is only 500 pc
in diameter).
In each case, we have corrected
the H$\alpha$ and H$\beta$ emission-line fluxes for the effects of stellar
absorption-lines
(using measures of the equivalent widths of the high-order
stellar Balmer absorption-lines in NGC 3256 and NGC 6240 and an assumed value
of 2 \AA\ for the other galaxies).
We have also corrected the data for foreground reddening
using the measured Galactic $HI$ column
density and assuming a standard extinction curve.

Figure 10 shows that not only do the extra-nuclear data points for these six
galaxies define a good correlation between
the amount of reddening and the depth of the $NaD$ line
along a given line-of-sight through the starburst and its outflow,
they define the {\it same} correlation
as that defined by the ensemble of all the $ISD$ nuclei in our sample.
The nuclear and off-nuclear points are
pretty well-mixed in Figure 10, although there is some tendency for the nuclear
lines-of-sight to have the larger values of reddening and deeper $NaD$
absorption-lines.
The correlation of $I_{5890}$ is better with the color of the stellar continuum
than with the Balmer decrement. This is reasonable because the 
$NaD$ line is observed in absorption against the background stellar continuum
(rather than against the emission-line gas)
and because the Balmer decrement is likely to be significantly affected
by dust directly associated with the emission-line gas itself
(in addition to the dust in the foreground material responsible for the $NaD$
absorption).

The observed Balmer decrements imply extinctions of $A_V$ $\sim$ 1 to 5
for a standard Galactic extinction curve.
Similar values are implied by the continuum colors: a typical starburst
is predicted (in the absence of reddening) to have a color of
$log[C_{65}/C_{48}] \sim$ -0.3 (Leitherer \& Heckman 1995), while the observed
colors in Figure 10 range from $log[C_{65}/C_{48}] \sim$ -0.2 to +0.3
(corresponding to $A_V$ = 0.7 to 4.2). Note also that in both Figure 10a and
10b, the extrapolation of the correlation to $I_{5890}$ = 1.0 (no
absorption, $C_f$ = 0) has an x-intercept at the intrinsic values expected
for an unreddened starburst
($log[C_{65}/C_{48}] \sim$ -0.3 and $log[H\alpha/H\beta]$
= 0.46).

In summary, the data imply that over regions with sizes of several or many kpc,
the outflows contain inhomogeneous {\it highly dusty} material. For a
standard Galactic extinction law and dust-to-gas ratio, the typical implied 
$HI$ columns are a few $\times$ 10$^{21}$ cm$^{-2}$. These $HI$ column densities
agree well with the estimates in section 3.4 above based upon the $NaI$
column density.

\subsection{Sizes, Masses, and Energies}

We have measured the size of the region over which
significantly blueshifted $NaD$ absorption is detected ($\Delta$$v$ $>$ 100
km s$^{-1}$) for the 12 outflow sources.
The sizes are listed in Table 4, and range from 1 to 10 kpc
in diameter. They
must be regarded as lower limits (since the background starlight usually
becomes too faint to detect the absorption at larger radii). Tracing the full
extent of the absorbing material farther out into the galactic halos will
probably require observing suitably bright background QSO's (see
Norman et al 1996).

These lower limits to the size of absorbing region can be used to estimate the
(minimum) mass and kinetic energy in the outflow. That is, for a region with
a surface area $A$, a column density $N_H$, and an outflow velocity
$\Delta$$v$:

\begin{equation}
M > 5 \times 10^8 (A/10 kpc^2) (N_H/3 \times 10^{21} cm^{-2})
M_{\odot}
\end{equation}

\begin{equation}
E > 2 \times 10^{56} (A/10 kpc^2) (N_H/3 \times 10^{21} cm^{-2}) (\Delta
v/200 km/s) erg
\end{equation}

We have scaled these relations using values for $A$, $N_H$, and
$\Delta$$v$ that are typical, and have assumed an equal contribution to $M$
and $E$ from the front (observed) and back sides of the outflow.

If we adopt a simple model of a constant-velocity, mass-conserving superwind
flowing
into a solid angle $\Omega_w$,  extending  to arbitrarily large radii from some
minimum radius ($r_*$ - taken to be the radius of the starburst within which
the flow originates), we obtain:

\begin{equation}
\dot{M} \sim 60 (r_*/kpc) (N_H/3 \times 10^{21} cm^{-2}) (\Delta v/200 km/s)
 (\Omega_w/4\pi) M_{\odot}/yr
\end{equation}

\begin{equation}
\dot{E}  \sim 8 \times 10^{41} (r_*/kpc) (N_H/3 \times 10^{21} cm^{-2})
(\Delta v/200 km/s)^3  (\Omega_w/4\pi) erg/s
\end{equation}

The statistics of the $ISD$ subsample in the present paper imply that
outflows are commonly observed in absorption in IR-selected starbursts
(12/18 cases). On the other hand, many of the outflow galaxies in the present
sample were selected from AHM on the basis of the strength of their $NaD$
line (objects above the $\sim$ 70th percentile in $EQ$).  If the presence of
observable blueshifted absorption is determined by viewing angle
(see section 3.3), this
suggests that $\Omega_w$/4$\pi$ lies in the range $\sim$ 0.2 to 0.6 (consistent
with the weakly-collimated bipolar outflows seen in well-studied
superwinds).

To put the above estimates into context, we can consider the rate at which
mass and energy are returned by massive stars. The median bolometric
luminosity of the 12 outflow galaxies in our sample is $L_{bol} \sim$ 2
$\times$ 10$^{11}$ L$_{\odot}$. The implied median rates of mass and kinetic
energy returned from supernovae and stellar winds are roughly
$\dot{M_{ret}}$ = 5
M$_{\odot}$ per year and
$\dot{E_{ret}}$ = 10$^{43}$ erg s$^{-1}$ respectively (e.g.
Leitherer \& Heckman 1995). Since $(\dot{M}/\dot{M_{ret}}) \sim$ 3 to 10,  the
absorption-line gas in the outflow must be primarily ambient gas that has
been loaded into the flow. This inference agrees with similar conclusions
about the hot X-ray-emitting gas in superwinds (section 4.2 below, and see
e.g. Strickland 1998;
Suchkov et al 1996; Heckman et al 1999).
Since $(\dot{E}/\dot{E_{ret}}) <$ 10\%, the absorbing gas does not
carry the bulk of the energy supplied by the starburst. Most of this energy
probably resides in the form of the thermal and kinetic energy of the much
hotter ($T > 10^{5.5}$ K) X-ray-emitting gas.

A bolometric luminosity of 2 $\times$ 10$^{11}$ L$_{\odot}$ corresponds
to a star-formation rate of about 12 M$_{\odot}$ per year (for a Salpeter IMF
extending from 1 to 100 M$_{\odot}$). Thus, the outflow rates estimated
from the $NaD$ lines are comparable to the star-formation rate: {\it the
feedback from massive stars drives the ejection of as much gas as is
being converted
into stars.} Similar inferences for starbursts have been made using the X-ray
and optical emission-line data (e.g. Suchkov et al 1996;
Heckman et al 1999; Della Ceca et al
1996,1999; Martin 1999).

\section{Discussion \& Implications}

\subsection{The Origin \& Dynamics of the Absorbing Material}

As discussed above, the red-most part of absorption-line profile in the
outflow objects  is close to $v_{sys}$, suggesting that absorbing material is
injected  from quiescent material at or near $v_{sys}$, and is then
accelerated up to some terminal velocity as it flows outward. This is
physically plausible, as the hot (X-ray emitting) outflowing gas interacts
hydrodynamically with colder denser material that is located either inside the
starburst, or in the inner portions of the galactic halo (see for example
Hartquist, Dyson, \& Williams 1997; Suchkov et al 1994; Strickland 1998).

Let us assume that a cloud of gas with a column density $N$, originally
located a distance $r_0$ from the starburst, is accelerated by a constant-
velocity superwind that carries an outward momentum flux $\dot{p}$ into a solid
angle $\Omega_w$. Ignoring the effects of gravity for the moment, the clump's
terminal velocity will be (Strel'nitskii \& Sunyaev 1973):

\begin{equation}
v_{term} = 420 (\dot{p}/7 \times 10^{34} dynes)^{1/2}(\Omega_w/1.6\pi)^{-
1/2} (r_{0}/kpc)^{-1/2}(N/3 \times 10^{21} cm^{-2})^{-1/2} km/s
\end{equation}

In this expression, we have used the momentum flux supplied by stellar
winds and supernovae (Leitherer \& Heckman 1995) in a starburst having a
bolometric luminosity equal to the median value for the outflow sample (2
$\times 10^{11}$ L$_{\odot}$) and have adopted the estimates in section 3
above for $N$ and $\Omega_w$. Starbursts with this luminosity have typical
estimated radii of roughly 1 kpc (see for example Heckman, Armus, \& Miley
1990; Meurer et al 1997).
From these elementary considerations, we conclude that the observed
terminal velocities (typically 400 to 600 km s$^{-1}$) are easily
accommodated.

Equation 7 also predicts that the outflow speeds will be larger in more
luminous starbursts. This trend is mitigated to some degree by the fact that
more powerful starbursts tend to be larger. Lehnert \& Heckman (1996b) and
Meurer et al (1997) argue that starbursts have a maximum characteristic
surface-brightness, which then implies $\dot{p} \propto L_{bol} \propto r^2$.
Together with Equation 7, this implies that such `maximum starbursts' will
have $v_{term} \propto L_{bol}^{1/4}$ (although the clouds can not be
accelerated to velocities larger than that of the flow that accelerates them!).
Our sample shows no convincing evidence of a trend for larger $v_{term}$
in the more luminous systems, but this sample covers a rather small range in
starburst luminosity. It will be instructive to extend
this study to dwarf starbursts with $L_{bol} < 10^9$ $L_{\odot}$.

Assume now that the cloud immersed in the outflow is subjected to a
gravitational force imposed by an isothermal  galaxy potential whose depth
corresponds  to a circular rotation speed $v_{rot}$. In order that the
outwardly-directed force due to the superwind exceed the inwardly-directed force
of gravity, the value for the cloud column density must satisfy the condition:

\begin{equation}
N < 7 \times 10^{21} (\dot{p}/ 7 \times 10^{34})(\Omega_w/1.6 \pi)^{-
1}(r/kpc)^{-1} (v_{rot}/200 km/s)^{-2} cm^{-2}
\end{equation}

Thus, the typical column densities estimated for the outflows ($\sim$ few
$\times$ 10$^{21}$ cm$^{-2}$) lie near the upper bound for material that
will flow out (rather than falling in). This may not be a coincidence: given a
range of cloud column densities, the blueshifted absorption-line will be
dominated by the largest-column-density clouds that can be expelled.
Alternatively, the observed column densities may simply arise because $N_H
\sim 2 \times$ 10$^{21}$ cm$^{-2}$ corresponds to a dust optical depth of 
unity in the continuum
at the wavelength of the $NaD$ doublet  (for a standard Galactic dust-
gas ratio). That is, continuum-emitting regions in these nuclei lying behind
sight-lines with much higher columns are invisible in optical light and
sources behind sight-lines with much lower columns contain little $NaI$.  It
would be interesting to test Equation 8  by measuring values of $N$ for
outflows in `low-intensity'
starbursts (small $\dot{p}/r$) and starbursts occurring in
dwarf galaxies (small $v_{rot}$). 

\subsection{Insights from Numerical Simulations}

The above elementary considerations give some simple physical insights
into the origin and dynamics of the absorbing material. More detailed insight
comes from hydrodynamical simulations of starburst-driven superwinds
(e.g. Tomisaka \& Bregman 1993, Suchkov et al 1994, Strickland
1998, Tenorio-Tagle \& Munoz-Tunon 1998). In these simulations the coolest
densest gas that has been
hydrodynamically disturbed
by the starburst is associated with the swept-up shell of
ISM that propagates laterally in the plane of the galaxy,
and fragments of the 
cap of the original superbubble shell now being carried vertically 
out of the disk by faster, more tenuous, wind material.
 
Shear between the hot shocked starburst ejecta
and the cool dense shell in the disk of the galaxy leads to entrainment
and stripping of cool dense gas into the wind flowing out of the
disk (through Kelvin-Helmholtz instabilities, and presumably additional
interchange processes such as thermal conduction and turbulent mixing
layers that can not be included in current simulations). Dense
gas already in the wind interior, for example the superbubble shell fragments,
is accelerated outward via the ram pressure of the
wind.

This process can be seen in Figure 11, in which the outward trajectories
of four typical entrained clouds are traced over an interval of 1.5 Myr
from a
2-D hydrodynamic simulation of M82's galactic wind (Strickland 1998). This is
based on the thick-disk ISM distribution of Tomisaka \& Bregman (1993).
In this model a mass of 10$^8$ M$_{\odot}$ is turned into stars
in an instantaneous burst (Salpeter IMF over the mass range 1 to 100
M$_{\odot}$). At a time 7 Myr after the burst, the resulting wind
has properties that are a reasonable match to M82. By this time, supernovae
and stellar winds have returned 1.3 $\times$ 10$^7$ M$_{\odot}$
and 6 $\times$ 10$^{56}$ ergs to the ISM.

Within $|z| \leq 1.5$ kpc of the disk there is $M = 1.9 
\times 10^{8} M_{\odot}$ of gas cooler than $T = 3 \times 10^{5}$ K. 
The majority of this gas is at the minimum temperature allowed 
in these simulations of $T \sim 6 \times 10^{4}$ K. This material
occupies a projected area of $\sim 2$ kpc$^{2}$, so the
average hydrogen column density of this gas is $N_{\rm H} \sim
9 \times 10^{21}$ cm$^{-2}$. This cool gas has a broad range of velocities,
from $v \sim 10$ -- $10^{3}$ km s$^{-1}$,
with a mode of $\sim 60$ km s$^{-1}$
(which is associated with the slow expansion of the outer shock in the
plane of the galaxy). The associated kinetic energy is 1.3 $\times$ 10$^{55}$
ergs. 
 
At higher distances above the disk there is much less cool gas.
For gas above $|z| = 1.5$ kpc the mass of cool gas, associated
primarily with the fragments of superbubble shell cap, is
$M = 1.5 \times 10^{7} M_{\odot}$. For a projected area of $\sim$
8.5 kpc$^{2}$, the average column density
is $N_{\rm H} \sim 1.6 \times 10^{20}$ cm$^{-2}$.
This cool gas high above the plane typically has higher
velocities than the gas within the plane of the galaxy, the
distribution of mass with velocity being approximately flat
between velocities of $v = 10^{2}$ -- $10^{3}$ km s$^{-1}$. 
The kinetic
energy associated with this gas is 1.7 $\times$ 10$^{55}$ ergs.

Thus, the total mass of cool gas in the wind is 2 $\times$ 10$^8$
M$_{\odot}$. This is twice as large as the mass of stars formed in
the burst, and is 16 times larger than the mass directly returned
by massive stars. In contrast, the total kinetic energy in the cool
gas (3 $\times$ 10$^{55}$) is only 5\% of the kinetic energy
returned by massive stars. The lion's share of this energy is the
form of thermal and kinetic energy of the hot ($T > 10^{5.5}$ K) 
gas in the wind.

It is worth noting several of the limitations of these simulations
with respect to their treatment of the cool dense ISM:
none of these simulations can explicitly include the cool dense clouds
of material within the ISM and starburst region that are known to
exist, and are thought to play a key role in ``mass-loading'' the outflow
(e.g. Hartquist, Dyson, \& Williams 1997). As a result, the cool dense
gas in these simulations is
confined to larger radii near the outer shock and to the shell fragments
(e.g. there is no way of explicitly treating the entrainment of clouds from 
within the starburst region itself). 
\footnote{The entrainment
of material into the hotter, more tenuous phases, from the hydrodynamical
destruction of such clouds can and has been simulated, but these
``mass-loaded'' simulations (Suchkov et al 1996, Strickland 1998)
do not consistently treat the properties of the clouds themselves.}
These simulations also have a minimum allowed gas temperature
of $T \sim 6 \times 10^{4} {\thinspace \rm K}$, due to the method of simulating
ISM turbulent pressure support by an enhanced thermal pressure.
Hence all the gas that would in reality have lower temperature is forced
to have this minimum temperature, which in turn affects the density
of this gas, and prevents us from knowing the exact distribution of this
mass between the gas phases cooler than this minimum temperature.
Similarly the processes of entrainment and acceleration of cool gas
into the wind are both uncertain and occur at (or below) the scale of
the physical resolution of these simulations.
 
Nevertheless, the results of these simulations are encouragingly
similar to the observed 
parameters in our sample of starburst outflows: the cool gas is predicted
to have column densities of several $\times 10^{20}$ to
$\sim 10^{22} \thinspace {\rm cm}^{-2}$, a mass that is comparable to
that of the stars formed in the burst, outflow velocities
in the range $v \sim 10^{2}$ -- $10^{3} {\thinspace \rm km} 
{\thinspace \rm s}^{-1}$, and a kinetic energy that is of-order 10$^{-1}$ of
the total kinetic energy returned to the ISM by the starburst. Since
the cool gas was originally cold dense material entrained and accelerated
by the hot outflow, the presence of substantial
amounts of dust
(section 3.5) is perhaps not too surprising.

\subsection{The Fate of the Outflow \& the Chemical Evolution of Galaxies}

As Figure 6 shows, the inferred terminal velocity in the outflows is typically
400 to 600 km s$^{-1}$, or about two to three times larger than the rotation
speed of the starburst's host galaxy. Are these velocities sufficient to expel
the gas from the galaxy altogether, or will the gas return to the galactic disk
as a fountain flow?

For an isothermal gravitational potential that extends to a maximum radius
$r_{max}$, and has a virial velocity $v_{rot}$, the escape velocity at a
radius $r$ is given by:

\begin{equation}
v_{esc} = [2v_{rot}(1 + ln(r_{max}/r))]^{1/2}
\end{equation}

Thus,  $v_{esc}$ = 3.0 $v_{rot}$  for  $(r_{max}/r)$ = 33 (e.g. $r$ = 3 kpc
and $r_{max}$ = 100 kpc).  As shown in Figure 6, the estimated terminal
velocities in the outflows are typically $v_{term} \sim$ 2 $v_{rot}$,
but $v_{term}$ is uncorrelated with $v_{rot}$. 

Similar results have been obtained for the hot X-ray-emitting gas in starburst
galaxies. This gas has temperatures of a few to ten million K in dwarf
galaxies (e.g. Della Ceca et al 1996; Strickland, Ponman, \& Stevens 1997),
$L_*$ disk
galaxies (e.g. Dahlem, Weaver, \& Heckman 1998; Read, Ponman, \& Strickland
1997),
and extremely powerful
starbursts in galactic mergers (e.g. Heckman et al 1999; Moran, Lehnert, \&
Helfand 1999; Read \& Ponman 1998). Martin (1999) has used these X-ray data to
estimate that the gas will escape from galaxies with
$v_{rot} <$ 130 km s$^{-1}$.

We can place these disparate data on common ground by comparing the
kinetic ($NaD$) and thermal (X-ray) energy per particle to the energy
needed for escape. For convenience, we do so by defining an energetically-
equivalent velocity for the X-ray gas. The terminal velocity in an adiabatic
superwind fed by gas at a temperature $T_X$ will be 
$v_X \sim (5kT_X/\mu)^{1/2}$,
where $\mu$ is the mean mass per particle (Chevalier \& Clegg 1985).
\footnote{
This is a conservative approach as it ignores any kinetic energy the
X-ray-emitting gas may already have. Currently the velocity 
and kinetic energy of the X-ray-emitting material 
in superwinds can not be measured directly, but 
numerical simulations suggest that the kinetic
energy of the hot gas is typically 2 to 3 times
its thermal energy (Strickland 1998).}

These results are shown in Figure
12, which includes the data from Fig. 6 plus 14
far-IR-bright galaxies for which analyses of broad-band ($\sim$ 0.1 to 10 keV)
X-ray data have been published. These are:
M82, NGC253, NGC3628, NGC3079, NGC4631 (Dahlem, Weaver, \& Heckman
1998), NGC1569 (Della Ceca et al. 1996), NGC1808 (Awaki et al. 1996), 
NGC2146 (Della Ceca et al. 1999), NGC3256 (Moran, Helfand, \& Lehnert
1999), NGC3310 (Zezas, Georgantopoulos, \& Ward 1998), NGC4038/4039 (
Sansom et al. 1996),
NGC4449 (Della Ceca, Griffiths, \& Heckman 1997),
NGC6240 (Iwasawa \& Comastri 1998), and Arp299 (Heckman et al 1999). 
In the cases where two-temperature plasma models were fit to the X-ray data,
we have plotted
both the corresponding outflow velocities.
The agreement between the two data sets is satisfactory.
There are three members of the $NaD$ outflow sample with X-ray data in
Figure 12, and the agreement between the $NaD$ terminal velocity and the X-
ray temperatures is reasonably  good: $v_{term}$ = 700 km s$^{-1}$ {\it vs.}
$v_X$ = 520 and 780 km s$^{-1}$ for NGC1808, $v_{term}$ = 580 km s$^{-
1}$ {\it vs.} $v_X$ = 490 and 800 km s$^{-1}$ for NGC3256, and $v_{term}$ =
580 km s$^{-1}$ {\it vs.} $v_X$ = 700 and 940 km s$^{-1}$ for NGC6240.
This suggests that
the fastest-moving $NaD$ absorbers are roughly co-moving with the hot
superwind fluid.

Figure 12 strongly suggests that shallower galaxy potential wells
will be less able to retain the newly-synthesized metals that are
returned to the ISM in the aftermath of a starburst. As has been
suggested many times (e.g. Wyse \& Silk 1985; Lynden-Bell 1992;
Kauffmann \& Charlot 1998) the
selective loss of metal-enriched gas from shallower potential
wells could explain both the mass-metallicity relation and radial
metallicity gradients in elliptical galaxies 
and galaxy bulges (Bender, Burstein,\& Faber
1993; Franx
\& Illingworth 1990; Carollo \& Danziger 1994;
Jablonka, Martin, \& Arimoto 1996; Pahre, de
Carvallo, \& Djorgovski 1998; Trager et al 1998).
 
A simple prediction of this idea would be that the relationship
between metallicity and escape velocity should saturate
(flatten) for the deepest potential wells - i.e. locations where
the local escape velocity exceeds the velocity of the outflowing
metal-enriched gas. Lynden-Bell (1992) has parameterized this in
a simple physically-motivated fashion by positing that the fraction of
metals produced by massive stars that
are retained by the galaxy ($f_{retained}$) is proportional to the
depth of the galaxy`s potential well ($\Phi$) for low-mass galaxies, but
asymtotes to $f_{retained}$ = 1 for the most massive galaxies. We
chose to cast his formulation as follows:

\begin{equation}
f_{retained} = v_{esc}^2/(v_{esc}^2 + v_{term}^2)
\end{equation}

Here $v_{term}$ is some characteristic velocity associated with the mixture
of supernova (and stellar wind) debris and entrained gas that is ejected
from the starburst.
It is assumed
to be a constant. For low-mass galaxies with $v_{esc} << v_{term}$,
$f_{retained} \propto v_{esc}^2 \propto \Phi$, 
or $f_{retained} \propto L_{gal}^{1/2}$
via the Faber-Jackson relation. Lynden-Bell shows that this simple
formula can reproduce the observed mass-metallicity relation for elliptical
galaxies over a range of $\sim$ 10$^6$ in galaxy mass,
and finds that the characteristic mass at which $f_{retained} = 1/2$
(e.g. a galaxy in which $v_{esc} = v_{term}$) corresponds to an elliptical
with $M_B$ $\sim$ -18 (adjusted to our assumed value of $H_0$ = 70). Such a
galaxy would have a line-of-sight velocity dispersion
$\sigma \sim$ 140 km s$^{-1}$,
corresponding to $v_{rot} = \sqrt{2} \sigma \sim$ 200 km s$^{-1}$
(Binney \& Tremaine 1987). Using equation 9 above, this would correspond
to $v_{esc} \sim$ 600 km s$^{-1}$. This in turn is a reassuringly good
match to the characteristic superwind outflow speeds implied by
Figure 12 ($\sim$ 400 to 800 km s$^{-1}$).

Thus, starburst-driven outflows might imprint a
relationship between metallicity and mass in ellipticals (and
bulges) over most of observed ranges for these two parameters.
While the loss of metal-enriched gas has the most
severe impact on dwarf elliptical galaxies, it may nevertheless have general
significance in galaxy chemical evolution.

\subsection{The Metal-Enrichment of the Intergalactic Medium}

The data discussed in this paper directly establish the flow of {\it metals}
out of highly-actively-star-forming galaxies
in the local universe, and the process is observed at high-redshift
as well (Franx et al 1997; Pettini et al 1998, 1999).
Such data allow us to estimate the column densities, outflow rates,
and outflow speeds of this material as a function of the rate of
star-formation. Meanwhile, over the past few years, the rate of high-mass
star-formation over the history of the universe has been measured
for the first time (e.g. Madau et al 1996; Steidel et al 1999;
Barger et al 1999).
This emboldens us to attempt to estimate
the amount of metals that have flowed out of galaxies,
thereby polluting the inter-galactic medium, over the course of cosmic time.

The discussion in section 3.4 above implies that gas is flowing out of
starbursts at a rate that is proportional to the rate of star-formation:
$\dot{M} = \alpha\dot{M_*}$ where $\alpha$ is one-to-a-few (see also
Martin 1999). The present-day mass in stars will be smaller than the total
mass turned into stars, since mass has been subsequently 
returned from these stars:
$M_{*,0} = \beta M_{*}$ where $\beta \sim$ 0.7 is reasonable for an old
present-day system like an elliptical or bulge. The discussion in section
4.3 implies that the outflowing gas will be mostly retained by galaxies with
the deepest potential wells, but mostly lost by the less massive systems.
Integrating equation 10 above over a Schechter luminosity function
implies that $\dot{M_{lost}} = \gamma\dot{M}$ with $\gamma \sim$ 0.5
(depending on the value of $v_{term}$ and the `mapping' of
$M_B$ to $v_{esc}$ in spheroids).

We then assume that over
cosmic time, we can attribute the construction of spheroidal systems
(elliptical and bulges) to starbursts (see Kormendy \& Sanders 1992;
Elmegreen 1999; Renzini 1999; Lilly et al. 1999).
If we further assume
that {\it all} star-formation in spheroids over the history of the universe
ejected
gas at the relative rate seen in local starbursts, then the ratio of the mass
of lost-gas to present-day stars in spheroids
would be $\sim \alpha\gamma/\beta$ (of-order unity). Fukugita, Hogan, \& Peebles
(1998 - hereafter FHP) estimate that the stars
in spheroidal systems today comprise $\Omega_{*,sph}$ = 2.6 $\times$ 10$^{-3}$
(for $H_0$ = 70). Since the
implied value for the gas expelled from forming spheroids is
comparable to this, this gas is therefore a significant repository
of baryons, but only of-order 10$^{-1}$ of the total estimated baryonic content
of the universe (FHP). In rich clusters,
nearly the entire stellar mass resides in spheroidal systems, while the
cluster potential well is deep enough to have retained all the mass
expelled by superwinds (e.g. Renzini 1997). The {\it observed} average ratio
of the mass of the intra-cluster medium to the stellar mass is $\sim$ 6
(FHP), so gas ejected by superwinds during spheroid formation would
comprise a significant, but minority share of this.

Equation 10 also 
implies that - integrated over the spheroid luminosity function - roughly
half of the metals produced by the stars will have been lost from the galaxies
and reside in the intergalactic medium or intracluster medium. Once
mixed with the metal-poor ``primordial'' baryons, the net metallicity
would be $\sim$ 1/6th solar in both the intracluster medium
and the general inter-galactic medium (assuming the
FHP global value 
$\Omega_{*,sph}/\Omega_{IGM} \sim$ 6, and assuming a mass-weighted mean
metallicity
equal to solar for stars in spheroids). The estimated metallicity
agrees reasonably well with the measured value of 0.3 solar in rich clusters 
(e.g. Renzini 1997). A measure of
the metal content of the present-day
general IGM may be possible with the next generation of UV and X-ray
space spectrographs
(e.g. Cen \& Ostriker 1999). 

These are not new arguments by any means (e.g. Gibson, Loewenstein, \&
Mushotzky 1997; Renzini 1997). What {\it is} new is that we are now
in a position to {\it observationally verify} that intense starbursts
of the kind that plausibly built galactic spheroids do indeed drive
mass and metals out at a rate and velocity perhaps high enough to
account for the observed inter-galactic metals.
The presence of such substantial amounts of inter-galactic
metals does not violate constraints imposed by the ``Madau diagram''
(star-formation rate {\it vs.} redshift), once reasonable corrections
for the effects of dust-extinction are made, nor does it violate
the limits set by the far-IR/sub-mm cosmic background (see for 
example Calzetti \& Heckman 1999; Renzini 1997).

\subsection{The Outflow of Dust}

The strong correlation
between reddening and the strength of the $NaD$ line in starbursts
(AHM; Veilleux et al 1995; Figure 10) implies that there is an intimate
relationship between the dust and gas, especially given the
close way in which the two track one another spatially
throughout the outflow (section 3.5, Figure 10, and see Phillips 1993
for the spectacular case of NGC 1808).
Moreover, we have argued above that
significant dust column densities in the absorbing matter are needed to shield
the $NaI$ from photoionization by the starburst's intense UV radiation field.

We therfore conclude that
{\it dust is being
expelled from starbursts at a significant rate.}
More quantitatively, for normal Galactic dust,
the observed reddening implies a dust surface mass density in the outflow
region of $\sim$10$^{-4}$ gm cm$^{-2}$, an outflowing dust mass
of $\sim$ 10$^6$ to 10$^7$ M$_{\odot}$ (see equation 3), and a dust
outflow rate of 0.1 to 1 M$_{\odot}$ yr$^{-1}$ (see equation 5).

Additional evidence for dusty galactic outflows comes from
a variety of observations. Spectroscopy with $HST$ and $HUT$ has
established that - just as in the case of the $NaD$ lines - the strong
$UV$ interstellar absorption-lines are frequently
blueshifted by several hundred km s$^{-1}$ in local starbursts
(Lequeux et al 1995; Heckman \& Leitherer 1997; Kunth et al 1998;
Gonzalez-Delgado et al 1998a).
Moreover, as discussed by Heckman et al (1998), there is
a strong correlation
between the equivalent widths of these $UV$ absorption-lines
and the reddening in the $UV$ that is analogous to the correlation between
reddening in the optical and the $NaD$ equivalents widths.
The $IUE$ spectra discussed by Heckman et al (1998)
do not resolve the $UV$ absorption-lines, and so can not verify
that the correlation is primarily driven by the covering fraction of the
absorbing dusty material (as in the case of the $NaD$ line). As
the archive of $HST$ $UV$ spectra of starbursts grows, it will be
possible to test this.

Images of several edge-on starburst and star-forming galaxies show
far-IR and/or sub-mm
emission extending one or two kpc along the galaxy minor axis
(Alton, Davies, \& Bianchi 1999; Alton et al 1998).
Multi-color optical images show that kpc-scale extraplanar dust filaments 
are common in star-forming edge-on
galaxies (Howk \& Savage 1997, 1999; Sofue, Wakamatsu, \& Malin 1994;
Phillips 1993; Ichikawa et al 1994). Imaging polarimetry
reveals light scattered by extraplanar dust in 
starburst galaxies (Scarrott, Eaton, \& Axon 1991; Scarrott et al. 1993;
Scarrott, Draper, \& Stockdale 1996;
Alton et al 1994; Draper et al 1995).
Zaritsky (1994) finds evidence
for very extended dust in the halos of spiral galaxies
based on the possible detection of reddening in background field galaxies.

As discussed by Howk \& Savage (1997) and Aguirre (1999b), there are a variety 
of mechanisms
by which an episode of intense star-formation could lead to the outflow
of dust grains. Radiation pressure can ``photo-levitate'' the grains
(Ferrara et al 1991; Ferrara 1998; Davies et al 1997),
the Parker instability could help loft
material out of the starburst disk (e.g. Kamaya et al 1996), or cold, dusty
gas in and around the starburst could be entrained and accelerated outward
by the hot outflowing X-ray gas in the superwind (Suchkov et al 1994, and
see section 4.2 above). 

The superwind mechanism is of
the most direct relevance to the present paper, so we briefly evaluate
its plausibility. First, we can show that
the outward force of the wind on even the largest grains will exceed
the inward force of gravity on the grain. For an isothermal galactic potential
this force-ratio at a distance $r$ from the starburst is given by:

\begin{equation}
F_w/F_g = 3 \dot{M}v_{term}/4\Omega_{w} r v_{rot}^2 a \rho
\end{equation}

where $a$ and $\rho$ are the radius and density of the grain (we take
$\rho$ = 2 gm cm$^{-3}$ as representative).
For the estimated properties of typical outflow sources in our sample
($\dot{M} \sim$ 25 M$_{\odot}$
per year, $v_{term} \sim$ 600 km s$^{-1}$, and $\Omega_{w}/4\pi \sim$ 0.4),
$F_w/F_g$ will be greater than unity for grains smaller than
7 $\mu$$m$ ($r$/10 kpc)$^{-1}$.

Next, we follow
Aguirre (1999b) and estimate the ratio of the sputtering
and outflow times for graphite grains
immersed in a hot galactic wind ($\tau_{sp}$/$\tau_{out}$).
For the typical parameters
we deduce for the outflows in our sample (see above),
this ratio is  $\tau_{sp}$/$\tau_{out}$ = 4 ($a$/0.1$\mu$$m$)($r$/10 kpc).
Thus, {\it large} grains could in fact survive the journey
to the galactic halo and beyond. The survivability of grains may actually be
higher than the above simple estimate if the grains are imbedded inside
cold gas clouds propelled by the hot outflow (so that the grains
are not directly exposed
to the hot gas).

{\it If} starburst and star-forming galaxies are indeed capable of ejecting
substantial quantities of dust, this could have a profound impact
on observational cosmology (e.g. Heisler \& Ostriker 1988; Davies et al 1997;
Ferrara 1998;
Ferrara et al 1999; Aguirre 1999a,b). However, to date, the
direct evidence for the existence of intergalactic dust is very sparse.
Thermal far-IR emission has been detected from the
ICM of the Coma cluster (Stickel et al 1998), and a
possible deficit of background QSO's seen through foreground
galaxy clusters has been reported (Romani \& Maoz 1992; but see Maoz 1995).

Aguirre (1999a,b) has recently calculated that a dusty inter-galactic medium
with $\Omega_{dust}$ = few $\times$ 10$^{-5}$ would have a visual
extinction ($\sim$ 0.5 magnitudes out to z = 0.7) that would be sufficient
to reconcile the Type Ia supernova Hubble diagram (Reiss et al 1998;
Perlmutter et al 1999) with a standard $\Omega_{M}$ = 1, $\Omega_{\Lambda}$
= 0 cosmology. Data on the optical colors of high-redshift supernovae
show no evidence for {\it reddening},
but Aguirre argues that intergalactic dust will have a much greyer
extinction curve than standard Galactic dust. This is plausible because
small grains will be more easily destroyed by sputtering 
during and after their journey into the IGM (see above).

In this context, it is instructive to estimate the cosmic mass
density of dust grains by the type of outflows investigated
in this paper. Aguirre (1999b) has considered this in more detail
from a somewhat different perspective, but comes to rather similar
conclusions.
Let us assume that superwinds associated with the
formation of galactic spheroids propelled dust and gas-phase metals
into the ICM and IGM, with an amount proportional to the mass
in the present-day stars in such systems. We further assume that the 
mass fractions
of the metals locked into grains in the ICM and IGM are $f_{g,icm}$
and $f_{g,igm}$ respectively. These assumptions imply:

\begin{equation}
\Omega_{dust,igm} = f_{g,igm}(1 - f_{g,icm})^{-1}\Omega_{spheroids}\Omega_{icm}
Z_{icm}\Omega_{stars,cl}^{-1}
\end{equation}

Following FHP and Renzini (1997), we take $Z_{icm}$ = 6.7 $\times$ 10$^{-3}$
(1/3 solar metallicity), $\Omega_{spheroids}$ = 0.0026 $h_{70}^{-1}$,
$\Omega_{icm}$ = 0.0026 $h_{70}^{-1.5}$, and $\Omega_{stars,cl}$ = 0.00043
$h_{70}^{-1}$. This implies $\Omega_{dust,igm}$ = 1.0$\times$ 10$^{-4}$
$f_{g,igm}(1 - f_{g,icm})^{-1}h_{70}^{-1.5}$. For a normal Galactic
dust/metals ratio ($f_g \sim$ 0.5), the implied value for $\Omega_{dust,igm}$
is twice as large as the value needed to explain the Type Ia
supernova-dimming (Aguirre 1999a).
Given the higher densities (and thus, faster grain sputtering times)
in the ICM compared to the IGM, we might expect
that $f_{g,icm} < f_{g,igm}$. More importantly, it
is also possible that $f_{g,igm} <<$ 0.5 due to
the destruction of dust in
superwinds and/or the 
IGM (but see Aguirre 1999b for an optimistic assessment).
While the above estimate for $\Omega_{dust,igm}$
should therefore probably be regarded as an absolute upper
bound, it is an intriguingly large one from a cosmological perspective.

Finally, we note that since intergalactic dust will emit as well
as absorb, its amount is constrained by
the cosmic background measured by $COBE$
(Ferrara et al 1999). Indeed, Aguirre \& Haiman (2000) argue that a significant
fraction of the detected cosmic far-IR and sub-mm background must have an
intergalactic origin if this dust is abundant enough to strongly
affect the Type Ia supernova Hubble Diagram.

\subsection{Relationship to ``Associated Absorption'' in AGN}

Over the past few years, it has become increasingly clear that a young
stellar population is present in the circumnuclear region of
a significant fraction of type 2 Seyfert galaxies (e.g. Heckman et al 1995,1997;
Gonzalez-Delgado et al 1998b;
 Schmitt et al 1999; Oliva et al. 1999). Most
recently, a near-UV spectroscopic survey of a complete sample of the brightest
type-2 Seyfert nuclei by Gonzalez-Delgado, Heckman, \& Leitherer
(2000) finds direct evidence for hot, young stars in roughly half
of the nuclei. 
In this paper we have established that starbursts drive outflows of cool
or warm gas with total column densities of a few $\times$ 10$^{21}$
cm$^{-2}$, velocities of a few hundred km s$^{-1}$, and covering factors
along the line-of-sight of typically 50\%. The implication then is that
this absorbing material should be detectable in those Seyfert nuclei
that also contain a circumnuclear starburst. 

In the standard ``unified''
scenario, type 1 and type 2 Seyfert nuclei are drawn from the same parent
population, with the former viewed from a direction near the polar axis
of an optically and geometrically-thick ``obscuring torus'' and the latter
from a direction near the equatorial plane of the torus (e.g. Antonucci
1993 and references therein). Thus, in type 1 Seyferts, any starburst-driven
outflow could be observed in absorption against the bright nuclear
continuum source. While the total column density of the
outflowing gas
should be similar to the flows studied in this paper, the
gas would be exposed to the intense ionizing continuum from the central
nucleus, and therefore would be significantly more highly-ionized. 

This can
be quantified as follows. The ionization state of photoionized gas is
determined by the ionization parameter:
\begin{equation}
U = Q/4\pi r^2 n c
\end{equation}
where $Q$ is the production rate of ionizing photons and 
$n$ is the electron density in the photoionized
material located a distance $r$ from the source. 
The radial density gradients observed in starburst-driven outflows
are consistent with predictions for clouds subjected to the
ram pressure associated with the superwind (Heckman, Armus, \&
Miley 1990): 
\begin{equation}
2n(r)kT \sim P(r) = \dot{p}/\Omega_w r^2
\end{equation}
where $\dot{p}$ is the rate at which the starburst feeds momentum into
the superwind.
For photoionized gas, $T \sim$ 10$^4$ K (e.g Osterbrock 1989), so
equations 13 and 14 together
imply that the magnitude of $U$ is set by
$Q/\dot{p}$,
and that $U$ will be independent of $r$ (neglecting
radiative transfer effects).
We adopt a generic Leitherer \& Heckman (1995) starburst model
(Salpeter IMF extending
up to 100 M$_{\odot}$ and a starburst lifetime of a few $\times$ 10$^7$
years),
include sources of ionization due
to both a starburst ($Q_*$) and the type 1 Seyfert nucleus ($Q_{sy1}$).
For a starburst and type 1 Seyfert nucleus of the same bolometric
luminosity, $Q_{sy1}/Q_*$ would be a factor of several.
We then obtain the following estimate for $U$:
\begin{equation}
U = 2.6 \times 10^{-3}(\Omega_w/4\pi)(1 + Q_{sy1}/Q_*)
\end{equation}
The predicted properties of the absorbing material then overlap
significantly with the ``associated absorbers'' seen in $UV$ spectra
of type 1 Seyfert nuclei (e.g. Crenshaw et al 1999; Kraemer et al 1999):
an incidence rate of roughly 50\%, a high line-of-sight covering
fraction, outflow velocities of 10$^2$ to 10$^3$ km s$^{-1}$, and
inferred ionization parameters of $\sim$ 10$^{-2}$. Crenshaw et al
(1999) find an essentially one-to-one correspondence between
the presence of $UV$ absorption-lines and soft X-ray absorption
by hotter and more highly ionized material (the ``warm absorber'').
We speculate that 
the hotter and more tenuous phases of the starburst superwind could
contribute to the warm absorber.

We emphasize that we are not proposing that all of the ``associated
absorption'' seen in type 1 Seyfert nuclei is produced by gas
in a starburst-driven outflow. In some cases, rapid variability
or the presence of absorption out of highly-excited lower levels
imply densities that are orders-of-magnitude higher than would be
tenable for material in a starburst-driven outflow (Crenshaw et al 1999
and references therein). However, it appears that the 
absorbing material in type 1 Seyfert nuclei can span a broad range
in physical and dynamical conditions (Kriss et al 2000).
Given important roles
for starbursts in the Seyfert phenomenon and for superwinds in the starburst
phenomenon, significant absorption due to the superwind material seems
unavoidable in some Seyfert nuclei.

\section{Conclusions}



We have discussed the results of moderate-resolution
($R$ = a few thousand) spectroscopy of 
the $NaI\lambda\lambda$5890,5896 ($NaD$) absorption-line in
a sample of 32 far-IR-selected starburst galaxies. These galaxies were
selected from either the far-IR-warm sample of Armus, Heckman, \&
Miley (1989) or the edge-on sample of Lehnert \& Heckman (1995),
and together span a range from 10$^{10}$ to few $\times$ 10$^{12}$
L$_{\odot}$ in IR luminosity. 
We found that the stellar contribution to the
$NaD$ absorption-line is negligible ($<$10\%) in some objects, but significant
($\sim$ 70\%) in others. We have thus divided our sample into
18 interstellar-dominated (``ISD'') objects ($<$ 30\% stellar
contribution) and 14 strong-stellar-contamination (``SSC'')
objects ($>$ 40\% stellar contribution).

The $NaD$ line lies within
70 km s$^{-1}$ of $v_{sys}$ in all the SSC objects
(consistent with a predominantly
stellar origin).
The $NaD$ lines in the SSC nuclei are about
0.2 dex narrower than expected for dynamics of the old stellar population
in the bulges of normal galaxies of similar disk rotation speed
and Hubble type. Thus,
dynamically ``cold'' material (red supergiants and/or
interstellar gas) in the inner part of the starburst makes
a significant contribution to the observed $NaD$ line in these
nuclei.

The kinematics of $NaD$ line are markedly different in the ISD objects.
The $NaD$ line is blueshifted by $\Delta$$v >$ 100
km s$^{-1}$ relative to the galaxy systemic velocity in 12 of the 18
cases (the
``outflow sources''), and the outflow can be mapped over a region
of a few-to-ten kpc in size.
In contrast, no objects in our sample
showed a net
{\it redshift} in $NaD$ of more than 100 km s$^{-1}$.
The outflow
sources are galaxies systematically viewed more nearly face-on than
the other galaxies in our sample:
69\% of the galaxies
with a ratio of semi-major to semi-minor axes $a/b \leq$ 2.0 show 
$NaD$ outflows,
while this is true for only 6\% of the flatter (more highly inclined)
galaxies. This is consistent with the absorbing
material being
accelerated out along the galaxy minor axis by a bipolar
superwind.
The absorbing
material typically spans the velocity range from near
the galaxy systemic velocity ($v_{sys}$) to a maximum blueshift
of 300 to 700 km s$^{-1}$. We therefore suggest that the outflowing
superwind ablates the
absorbing gas from ambient clouds at $\sim v_{sys}$, and then
accelerates it up to a terminal velocity similar to the wind speed. We
found no correlation between the widths of the H$\alpha$
emission-line and the $NaD$ absorption-line
subsamples. Evidently, the dynamics of the more tenuous absorbing
gas is largely decoupled from that of the dense
(high emission-measure) gas that
provides most of the nuclear line-emission.

The ratio of the equivalent widths of the two members of the $NaD$
doublet ($R$) ranges from 1.1 to 1.7 in the ISD sample, implying
that the doublet is optically-thick. However,
$R$ does not correlate with the residual relative intensity
at the ``bottom'' of the stronger $\lambda$5890 line profile 
($I_{5890}$), which ranges from 0.14 (nearly black) to 0.7.
Thus, the optically-thick gas does not fully cover
the emitting stars (covering factor $\sim$ 1 - $I_{5890}$). The observed
equivalent width of the $NaD$ line is then set by the product
of velocity dispersion and covering factor for the
absorbing gas, and we showed that the
latter quantity is the dominant one.
Using two variants of the classic doublet-ratio technique, we estimated that
the $NaI$ column densities are $logN_{NaI}$ = 13.5 to 14 cm$^{-2}$. This
is roughly consistent with column densities measured in a few cases for $KI$
using the optically-thin $\lambda$$\lambda$7665,7699 \AA\ doublet
(assuming a solar $NaI/KI$ ratio). The total gas columns are uncertain,
but the empirical correlation between $N_{NaI}$ and $N_H$ in the
ISM of the Milky Way implies $N_H \sim$ few $\times$ 10$^{21}$ cm$^{-2}$.

We found a strong correlation in the ISD sample between the reddening of the
observed stellar
continuum and the depth of the $NaD$ absorption-line, and a significant
but weaker correlation of the line-depth with the reddening of the
Balmer emission-lines. Evidently,
the gas responsible for the $NaD$ absorption is very dusty. The typical
implied reddening is $E(B-V) \sim$ 0.3 to 1 magnitudes over regions
several-to-ten
kpc in size. For a normal dust-to-gas ratio, the corresponding column
densities are $N_H \sim$ few $\times$ 10$^{21}$ cm$^{-2}$ (in agreement
with the above estimate).

The inferred column densities and measured outflow velocities and
sizes imply that the typical mass and kinetic energy associated
with the absorbing gas is of-order 10$^9$ M$_{\odot}$ and
10$^{56}$ erg, respectively. The estimated outflow rates of mass and energy
are typically 10 to 100 M$_{\odot}$ per year and
10$^{41}$ to 10$^{42}$ erg s$^{-1}$. The mass outflow rates are comparable
to the estimated star-formation rate, and much larger than the rate at
which massive stars are returning mass to the ISM. Thus, powerful
starbursts can eject as much gas as is being converted into stars,
and most of this gas is ambient material that has been ``mass-loaded''
into the hot gas returned directly by supernovae and stellar winds.
The energy outflow rates in the absorption-line gas are of-order 10$^{-1}$ of
the rate at which
massive stars supply mechanical energy. Most of the energy
returned by massive stars probably
resides in the kinetic and thermal energy of the much hotter X-ray-emitting
gas. We showed
that the overall properties of the absorbing gas in the outflow sources
can be easily reproduced in the context of simple analytic
estimates for the properties of interstellar clouds accelerated
by the ram pressure of the hot high-speed wind seen via its X-ray emission.
Detailed hydrodynamical simulations of galactic winds, while still missing some
essential physics, also predict the observed properties of the
cool absorbing gas.

We have discussed the implications of our results for the chemical
evolution of galaxies and the intergalactic medium.
The estimates derived for $v_{term}$ using the $NaD$ line in the outflow
sources agree reasonably well with the outflow speeds implied for
an adiabatic wind ``fed'' by hot gas whose temperature is measured
by the observed X-ray-emitting gas. The typical implied values
are 300 to 800 km s$^{-1}$, and are independent of the rotation speed
of the ``host galaxy'' over the range $v_{rot}$ = 30 to 300 km s$^{-1}$,
confirming and extending the result in Martin (1999) based on X-ray data alone.
This strongly suggests that the outflows selectively escape the potential
wells of the less massive galaxies.
We considered a simple model based on Lynden-Bell (1992) in which the fraction
of starburst-produced metals
that are retained by a galaxy experiencing an outflow is proportional to the
galaxy potential-well
depth for galaxies with $v_{esc} < v_{term}$, and asymtotes to
full retention for the most massive galaxies ($v_{esc} > v_{term}$).
For $v_{term}$ in the range we measure, such a simple prescription
can reproduce the observed mass-metallicity relation for elliptical
galaxies and deposit the required amount of observed metals in the
intra-cluster medium. If the ratio of ejected metals to stellar
spheroid mass is the same globally as in clusters of galaxies,
we predicted that the present-day
mass-weighted  metallicity of an intergalactic medium
with $\Omega_{igm}$ = 0.015 will be $\sim$ 1/6 solar (see also
Renzini 1997).

We have summarized the evidence that starbursts are ejecting significant
quantities
of dust, emphasizing the results from the present paper. {\it If}
this dust can survive a trip into the intergalactic medium and remain
intact for a Hubble time,
we estimated that the upper bound on the global amount of intergalactic
dust is $\Omega_{dust}$ $\sim$ 10$^{-4}$. While this is clearly an
upper limit, it is a cosmologically interesting one: 
Aguirre (1999a,b) argues that dust this abundant
could in principle obviate the need for a positive cosmological
constant, based on the Type Ia supernova Hubble diagram.

Finally, given the mounting evidence for a connection between starbursts and
the Seyfert phenomenon, we have suggested that outflows like those studied
here may account for some (but not all) aspects of the ``associated
absorption'' seen in type 1 Seyfert nuclei.

\acknowledgements

We would like to thank Ken Sembach for useful on-going discussions and advice.
Discussions with David Neufeld, Mark Voit, Don York, and Donna Womble were
helpful during the formative stages of the project. The partial support
of this project by NASA grant NAGW-3138 is acknowledged.


\include{tab1}

Note. Col. (2) --- Galaxy systemic velocity (km s$^{-1}$) in the heliocentric
frame.  In order of preference, these are determined from:  galaxy
rotation curves (r), global $CO$ 115 GHz emission-line profiles (c), nuclear
stellar velocities (s), global $HI\lambda$21cm emission-line profiles (h),
and nuclear optical emission-line profiles (e).  The rotation curve velocities
(r) are taken from LH95 except for NGC2146 from Prada et al (1994).
Items marked `n' come from NED.  Items marked `e' or `s' are based on
data obtained during the observing runs discussed in the present paper.
Items marked `c' are:  NGC 660 (Elfhag et al 1996; Young et al 1995),
NGC 1614 (Elfhag et al 1996; Young et al 1995; Aalto et al 1991;
Sanders, Scoville, \& Soifer 1991; Casoli et al 1991), M 82 (Lo et al
1987), IRAS10173+0828 (Planesas, Mirabel, \& Sanders 1991), NGC 3256
(Aalto et al 1991; Casoli et al 1991; Mirabel et al 1990),
IRAS10565+2448 (Downes \& Solomon 1998), and Arp 220 (Young et al 1995;
Solomon, Downes, \& Radford 1992).  Based on the intercomparison
of independent measurements for a given galaxy, the typical uncertainties
in $v_{sys}$ range from 10 km s$^{-1}$ for the nearby, relatively
normal galaxies
to as much as 100 km s$^{-1}$ for the most distant systems (generally,
highly disturbed mergers).
Col. (3) --- Total infrared luminosity from 8 to 100 microns, based on IRAS
data and the definition of L$_{IR}$ given in Sanders \& Mirabel (1996).  We
assume throughout that $H_0$ = 70 km s$^{-1}$ Mpc$^{-1}$.
Col. (4) --- Blue absolute magnitude for the galaxy, corrected for
foreground (Galactic) extinction, but not for internal extinction.
Taken from LH95 when available (adjusted to $H_0$ = 70 km s$^{-1}$
Mpc$^{-1}$) or based on the data in NED or Armus, Heckman, \& Miley
(1987).
Col. (5) --- The ratio of the optical semi-major to semi-minor axes.  These
are taken (in order of preference) from LH95, the images published in
Armus, Heckman, \& Miley (1987; 1990), or NED.  For the highly disturbed
merging systems, we have measured this ratio at intermediate radii
(excluding both faint tidal tails and the inner regions where dust
obscuration is most significant).
Col. (6) --- The amplitude of the rotation speed of the galaxy.  In order
of preference, we have based these on rotation curves (``r'' from LH95),
global $HI\lambda$21cm profiles corrected for inclination and turbulence
(``h'' - see LH95 for details), and global $CO$ 115 GHz emission-line
profiles using the half-width at 20\% of the peak intensity and then
correcting for inclination (``c'' - using the same data as in Column 2).
For M 82 we have replaced the value listed in LH95 by the more recent
determination by Sofue (1998).
The uncertainties in most cases are dominated by the inclination correction.
We estimate the resulting uncertainties to be $<$ 0.1 dex for all but the
cases of mergers and strongly interacting galaxies, where the
inclination corrections lead to an uncertainty of roughly 0.2 dex
(denoted by :).
Col. (7) --- Sample from which the galaxy was drawn (Armus, Heckman, \& Miley
1989; Lehnert \& Heckman 1995).
Col. (8) --- Observing runs used in this paper (see Table 2).

\include{tab2}

\include{tab3}
Note. Col. (2) --- The estimated contribution to the observed $NaD$ line by
cool stars (the remainder is interstellar in origin). See text for
details. Galaxies with $f_* \geq$ 40\% are members of the
strong-stellar-contamination sample (SSC), while those with
$f_* \leq$ 30\%  are members of the interstellar-dominated
(ISD) sample. Based on the agreement between $f_*$ determined by the two
independent techniques discussed in the text, the uncertainty is
typically $\pm$10\% .
Col. (3) --- Heliocentric velocity of the $NaD$ absorption-line (km s$^{-1}$)
measured by fitting the profile with a pair of Gaussians constrained
to have the separation in wavelength appropriate for the red-shifted
$NaD$ doublet. Based on comparison of independent measurements of this
quantity in the cases for which we have multiple spectra,
we estimate that the typical measurement uncertainty
is $\pm$ 20 km s$^{-1}$. 
Col. (4) --- The velocity difference between the $NaD$ absorption-line
and the galaxy systemic velocity in the galaxy rest-frame:
$\Delta$v = (v$_{NaD}$ - v$_{sys}$)/(1 + v$_{sys}$/c). The relevant
quantities are given in Col.2  of Table 1 and Col. 3 of this Table.
A typical uncertainty in this velocity difference is $\pm$ 20 km s$^{-1}$,
for the relatively bright nearby galaxies with well-determined values
for $v_{sys}$ up to $\pm$ 100 km s$^{-1}$ for the most distant and
far-IR-luminous
galaxies (highly disturbed mergers with uncertain $v_{sys}$).
Col. (5) --- The full-width at half-maximum of each of the two Gaussians
fit to the doublet (km s$^{-1}$).
$W$ was constrained to be the same for
the two doublet members. The listed value has had the instrumental
contribution to the measured value removed by assuming the intrinsic
and instrumental widths add in quadrature: W = [W$_{obs}^2$ -
W$_{instr}^2$]$^{1/2}$. Based on comparison of independent measurements of this
quantity in the cases for which we have multiple spectra,
we estimate that the typical measurement uncertainty
is $\pm$ 20 km s$^{-1}$. 
Col. (6) --- The rest-frame equivalent width (\AA) for the $NaD$ doublet.
Based on comparison of independent measurements of this
quantity in the cases for which we have multiple spectra,
we estimate that the typical measurement uncertainty
is $\pm$ 0.2 \AA.
Col. (7) --- The normalized residual intensity at the center of the $NaD$
$\lambda$5890 line profile (I$_{5890}$ = 0 corresponds to a totally black
line center). This has been corrected for the effect of the spectral
resolution assuming Gaussian profiles: (1-I$_{5890}$) = (W$_{obs}$/W)(1 -
I$_{5890,obs}$). Based on comparison of independent measurements of this
quantity in the cases for which we have multiple spectra,
we estimate that the typical measurement uncertainty
is $\pm$ 0.02.
Col. (8) --- The ratio of the equivalent widths of the
$NaD\lambda\lambda$5890,5896 transitions. A ratio $R$ = 2 (1) corresponds
to an optical depth of 0 (infinity). Based on comparison of independent
measurements of this
quantity in the cases for which we have multiple spectra,
we estimate that the typical measurement uncertainty
is $\pm$ 0.1.
Col. (9) --- The full-width at half-maximum of a Gaussian fit to the nuclear
$H\alpha$ emission-line. These are taken from AHM, LH95, or our own
unpublished spectra. The listed value has had the instrumental
contribution to the measured value removed by assuming the intrinsic
and instrumental widths add in quadrature. Typical uncertainties
are $\pm$ 20 km s$^{-1}$.

\include{tab4}
Note. Col. (2) --- The ratio of the nuclear H$\alpha$ and H$\beta$ emission-line
fluxes. These have been corrected for the effects of underlying stellar
absorption-lines
(assuming a stellar equivalent width of 1.5 \AA) and for foreground
Galactic extinction (using a standard Galactic extinction curve and
the measured Galactic $HI$ column density. The data come from AHM,
Veilluex et al (1995), data from runs 3 or 6 (Table 2), Vaceli et al
(1997), or Dahari \& DeRobertis (1988). Typical uncertainties are
$\pm$5\%. Unreddened ionized gas
would have a flux ratio of 2.86 for standard Case B conditions.
Col. (3) --- The ratio of the flux densities (F$_{\lambda}$) in the nuclear
continuum
near the wavelengths of H$\alpha$ and H$\beta$. The values have
been corrected for foreground Galactic extinction (see above).
The data come from AHM, Veilluex et al (1995), or our runs 3 or 6 (Table 2).
Typical uncertainties are $\pm$5\%.
Note that an unreddened starburst corresponding to constant star-formation
for 30 Myr would have an intrinsic color in these units of 0.5.
Col. (4) --- The projected size (in kpc) of the region along the spectrograph
slit exhibiting strongly blueshifted (by $>$ 100 km/s) $NaD$
absorption. In NGC1572, NGC1614, NGC3256, NGC7552, and NGC7582
we have measured this along two position angles.
Col. (5) --- An estimate of the terminal velocity implied by the $NaD$
absorption-line profile (v$_{term}$ = $\Delta$ v + 0.5 W). See Table 3.
Col. (6) --- The rotation speed of the starburst galaxy. See Table 1.

\newpage

\figcaption [] {a) through d)
Normalized spectra of the nuclear $NaD$ absorption-line
profile in the 32 galaxies in our sample. The region displayed is
typically 2 by 4 arcsec in size
centered on the region of peak brightness in the red continuum.
Each displayed spectrum covers an observed range of 60 \AA\
($\sim$ 3000 km s$^{-1}$). Note the weak foreground Galactic $NaD$
absorption at 5890,5896 \AA\ in NGC660, NGC2146, and NGC4945.}

\figcaption [] {Histogram of the difference between the mean radial velocity
of the $NaD$ line and the galaxy systemic velocity corrected to the
galaxy rest-frame - see Table 3. The $ISD$ (interstellar-dominated)
lines are indicated by vertical hashing and the $SSC$ (strong-stellar
contamination) lines by horizontal hashing. The majority of the $ISD$
lines are blueshifted by at least 100 km s$^{-1}$, while the $SSC$
lines are all close to $v_{sys}$. Uncertainties in $\Delta$$v$
range from $\pm$20 to 100 km s$^{-1}$ - see text and Table 3.
A typical uncertainty is represented by the plotted error-bar.}

\figcaption [] {Histogram of the full-width at half-maximum of each
of the two members of the $NaD$ doublet, corrected for the effects
of instrumental resolution and in the galaxy rest-frame. See Table 3.
The $ISD$ (interstellar-dominated)
lines are indicated by vertical hashing and the $SSC$ (strong-stellar
contamination) lines by horizontal hashing. The $ISD$ lines are much
broader than the $SSC$ lines. Typical uncertainties in $W$ are
$\pm$20 km s$^{-1}$, as indicated by the plotted error-bar.}

\figcaption [] {Plot of the full-width at half-maximum ($W$) {\it vs.}
the blueshift of the $NaD$ doublet ($\Delta$$v$) for the sources
showing nuclear outflows ($\Delta$$v \leq$ 100 km s$^{-1}$). This
Figure omits IRAS11119+3257, which has a blueshift of $\sim$ 10$^3$
km s$^{-1}$. The diagonal line shows the relation
$W$ = 2 $\Delta$$v$, expected in the case that gas is injected
into the outflow at $v \sim v_{sys}$, and accelerated up to
a terminal velocity $v_{term} \sim \Delta$$v$ + 0.5$W$.
Uncertainties in $\Delta$$v$
range from $\pm$20 to 100 km s$^{-1}$, and typical uncertainties in $W$ are
$\pm$20 km s$^{-1}$ (as indicated by the plotted error-bar).}

\figcaption [] {Plot of the galaxy rotation speed (Table 1)
{\it vs.} the full-width at half maximum of the $NaD$ line
(Table 3). The $ISD$ (interstellar-dominated)
lines are indicated by solid dots and the $SSC$ (strong-stellar
contamination) lines by hollow dots. The empirical relations between
bulge stellar velocity dispersion and disk rotation speed
found by Whittle (1992) for normal Sa (Sc) spiral galaxies are indicated
by the lower (upper) diagonal line. The $ISD$ lines are nearly all broader
than these relations, while the $SSC$ lines are nearly all narrower.
Typical uncertainties are $\pm$20 km s$^{-1}$ for $W$ and range
from $<$ 0.1 dex to 0.2 dex for $v_{rot}$ (dominated by the uncertain
inclination correction in highly disturbed systems). Typical uncertainties
are indicated by the plotted error-bar.}

\figcaption [] {Plot of the galaxy rotation speed (Tables 1 and 4)
{\it vs.} the inferred terminal velocity of the outflow
($v_{term} = \Delta$$v$ + 0.5$W$;  see Table 4) for the sources
showing outflows in $NaD$ line ($\Delta$$v \leq$ -100 km s$^{-1}$).
The diagonal line shows the relation
$v_{term}$ = 2$v_{rot}$. The terminal velocity shows no dependence
on the rotation speed. Uncertainties in $v_{term}$
range from $\pm$20 to 100 km s$^{-1}$, and 
from $<$ 0.1 dex to 0.2 dex for $v_{rot}$ (dominated by the uncertain
inclination correction in highly disturbed systems). The typical 
uncertainties are indicated by the plotted error-bar.}

\figcaption [] {Plot of the
normalized residual intensity at the center of the $\lambda$5890
transition ($I_{5890}$) {\it vs.}
the ratio of the 
equivalent widths of the $\lambda$5890
and $\lambda$5896 members of the $NaD$ doublet ($R$),
for the $ISD$ (interstellar-dominated)
$NaD$ lines. See Table 3. $R$ = 1 (2) corresponds to the limit
of optically-thick (-thin) conditions. There is no correlation,
implying that $I_{5890}$ is determined primarily by covering factor
rather than optical depth. Typical uncertainties are $\pm$ 
0.02 in $I_{5890}$ and $\pm$ 0.1 in $R$, as shown by the plotted
error-bar.}

\figcaption [] {Plot of the normalized residual intensity at the 
center of the $\lambda$5890
transition ($I_{5890}$) {\it vs.} the rest-frame equivalent width of the
$NaD$ doublet for the $ISD$ (interstellar-dominated)
lines. There is a significant correlation, implying that
the equivalent width depends on the covering fraction of the
optically-thick absorbing gas. Typical uncertainties are $\pm$
0.02 in $I_{5890}$ and $\pm$ 0.2 \AA\ in $EQW_{NaD}$, and are shown
by the plotted error-bar.}

\figcaption [] {Plot of the full-width-at-half-maximum {\it vs.}
the rest-frame equivalent width of the
$NaD$ doublet for the $ISD$ (interstellar-dominated)
lines. There is no correlation, implying that
the equivalent width does not depend on the velocity dispersion of the
optically-thick absorbing gas. Typical uncertainties are $\pm$
20 km s$^{-1}$ for $W$ and $\pm$ 0.2 \AA\ in $EQW_{NaD}$, as shown
by the plotted error-bar.}

\figcaption [] {a) Plot of the normalized residual intensity at the 
center of the $\lambda$5890
transition ($I_{5890}$) {\it vs.} the log of the color of the optical continuum
(the ratio of $F_{\lambda}$ at rest wavelengths of 6560 and 4860
\AA). Points plotted as solid dots are the nuclei of the $ISD$ 
(interstellar-dominated) sample members (Table 4). Other points
are off-nuclear locations in M82, NGC3256, NGC6240, Mrk273, IRAS03514+1546,
and IR10565+2448.
The deeper the
$NaD$ line (higher covering factor), the more-reddened the background starlight.
The correlation is obeyed by both the nuclear and off-nuclear regions.
An unreddened starburst population should have $log(C_{65}/C_{48})$
= -0.3. For a standard Galactic reddening curve, the implied
$A_V$ ranges up to roughly 4 magnitudes for the most-reddened sight-lines.
Typical uncertainties for the nuclear (extra-nuclear) data are indicated
by the error-bar in the lower-left (upper-right) of the plot.
b) As in a), except that $I_{5890}$ is plotted {\it vs.}
the log of the Balmer decrement (H$\alpha$/H$\beta$ flux ratio). Again, the
more-reddened sight-lines correspond to the deepest $NaD$ line profiles
(highest covering factors). The log of the intrinsic H$\alpha$/H$\beta$ flux
ratio is 0.46, and the implied values of $A_V$ range up to $\sim$ 5 magnitudes
for the most-reddened sight-lines.} 

\figcaption [] {Logarithm of the gas number density (in units of 
${\rm cm}^{-3}$)
in the hydrodynamic simulation described in the text, at four
different epochs. The figure shows dense cool gas, either entrained
into the flow at the walls of the cavity or remnants of the fragmented
superbubble shell, being swept up and carried out of the galaxy by the wind. 
As material is locally in pressure equilibrium,
the densest material visible is also the coolest, and might be
considered analogous to the dense cool gas responsible for the
optical absorption lines. 
The four clumps or clouds shown have average velocities (over the 1.5 Myr
period shown) of 177 km s$^{-1}$ (cloud A), 
540 km s$^{-1}$ (cloud B), 348 km s$^{-1}$ (cloud C) and
858 km s$^{-1}$ (cloud D) respectively. }

\figcaption [] {As in Figure 6, except that we have added galaxies
in which we have estimated wind outflow velocities from the
observed temperature of the hot X-ray-emitting gas via the
relation from Chevalier \& Clegg (1985): $v_{term} \sim (5kT_X/\mu)^{1/2}$.
The X-ray temperatures are taken from the references listed in the text.
The data points
based on the $NaD$
profile are indicated by solid dots and the points based on the X-ray data
are indicated by hollow dots. Note that the two data sets are consistent with
each other,
imply that the outflow speed is independent of the host galaxy
potential well depth, and thus suggest that outflows will
preferentially escape from the least massive galaxies. The two diagonal
lines indicate the galaxy escape velocity under the assumption
that $v_{esc}$ = 2 $v_{rot}$ and $v_{esc}$ = 3 $v_{rot}$ respectively
(see equation 9 in the text). Typical uncertainties in the
X-ray ($NaD$) estimates of $v_{term}$ are shown by the error-bar on the
bottom right (upper center).}

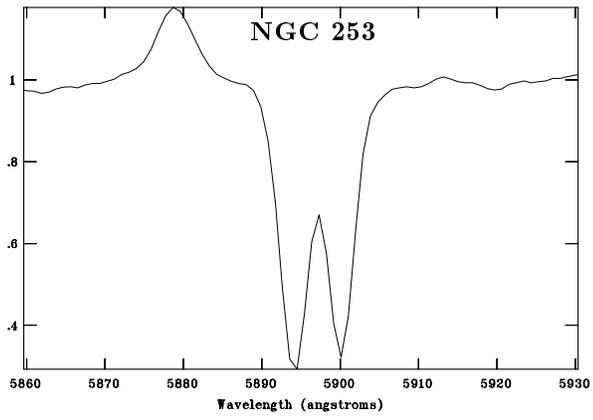
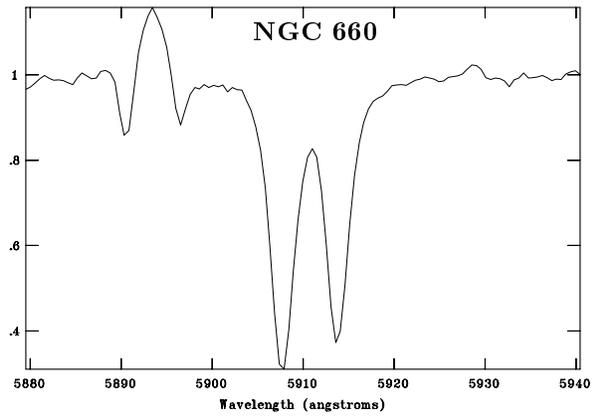
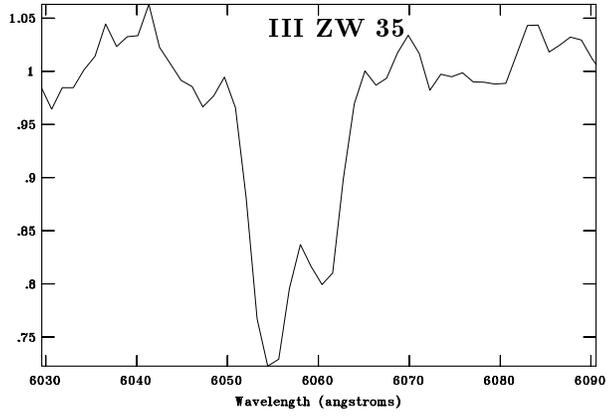
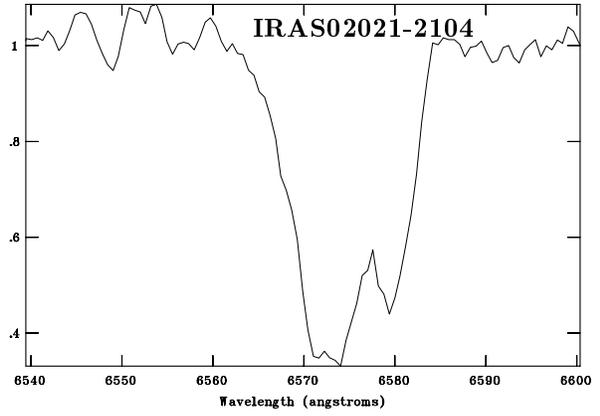
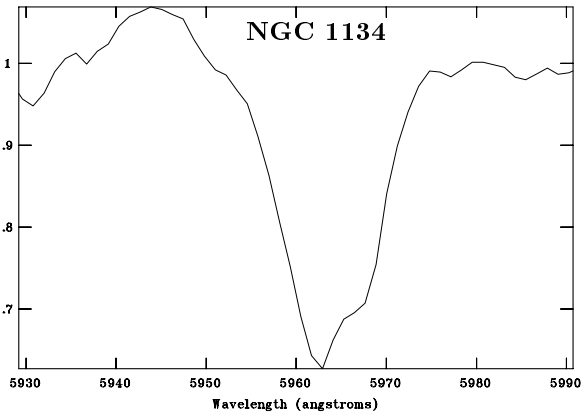
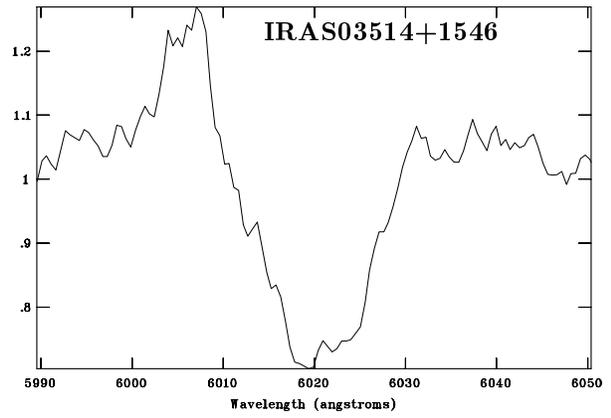
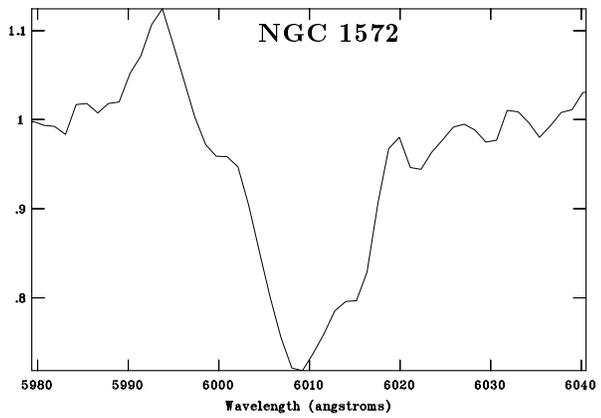
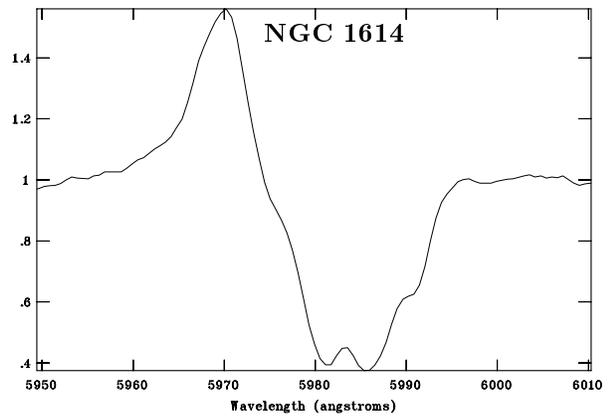
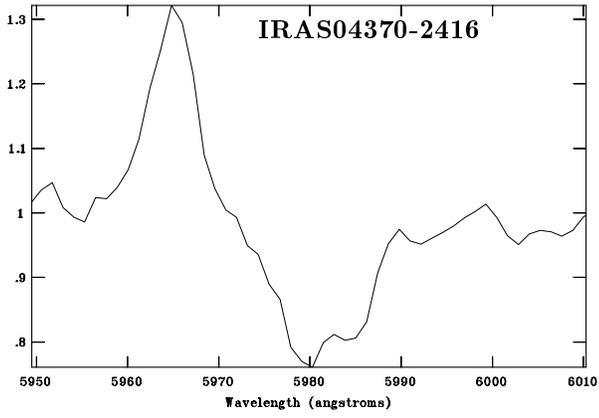
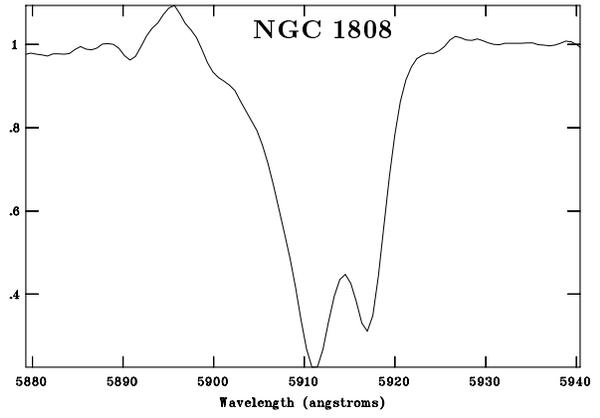
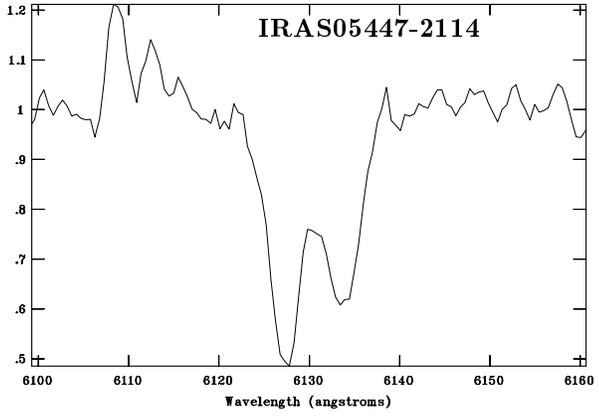
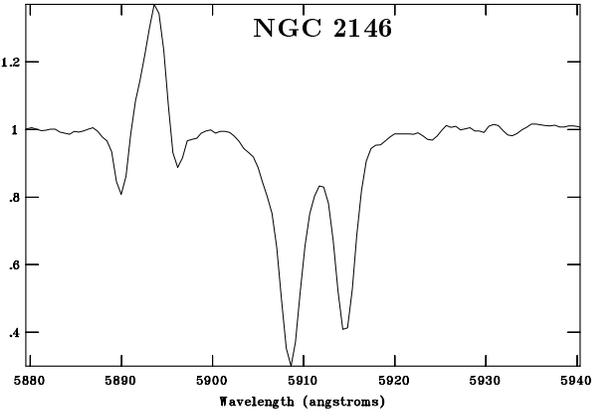
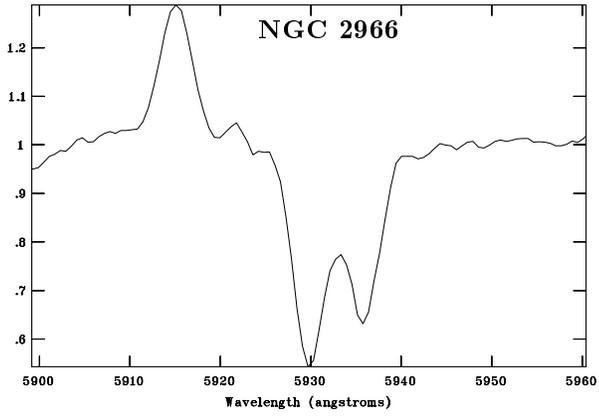
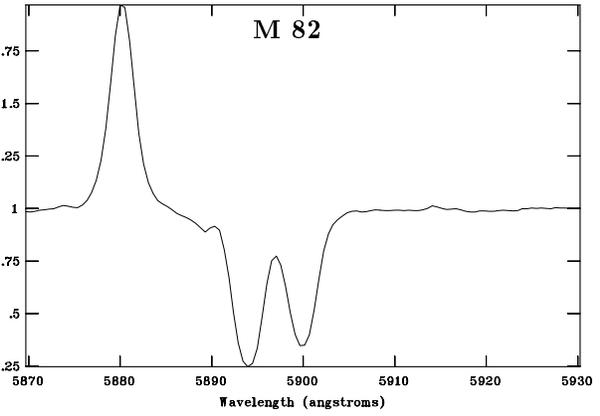
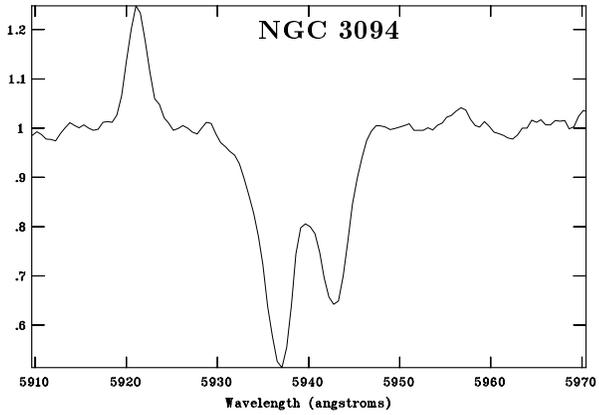
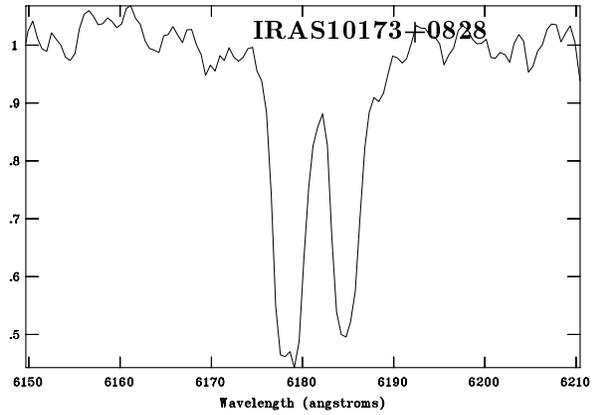
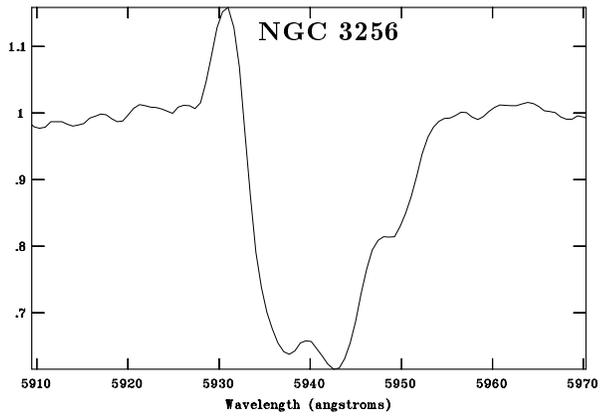
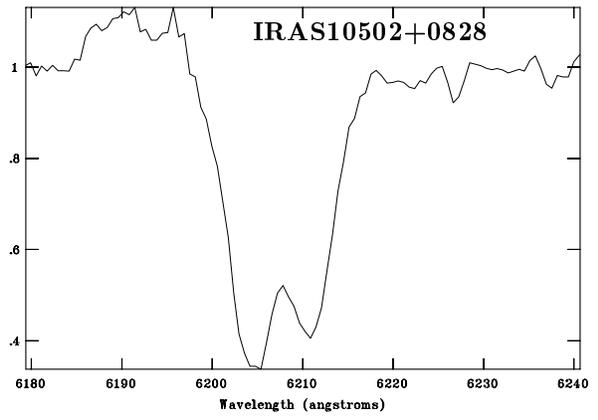
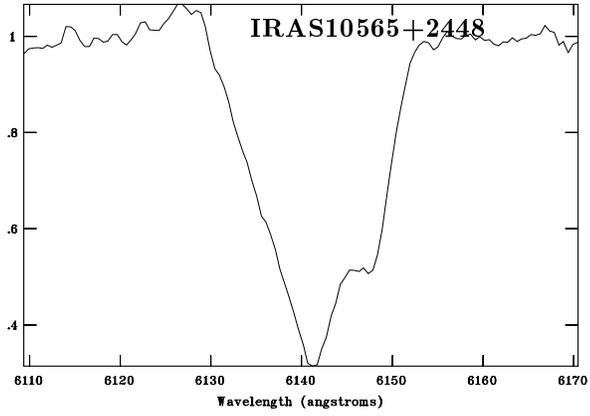
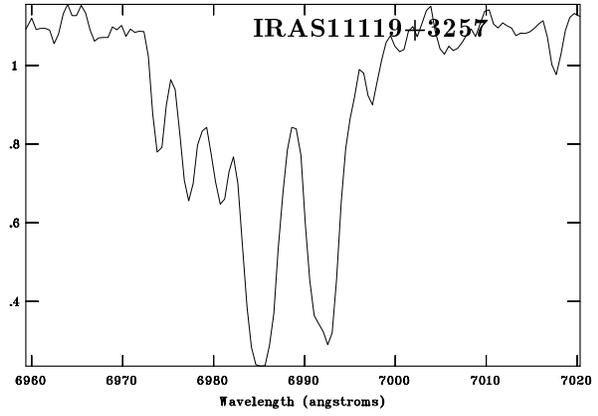
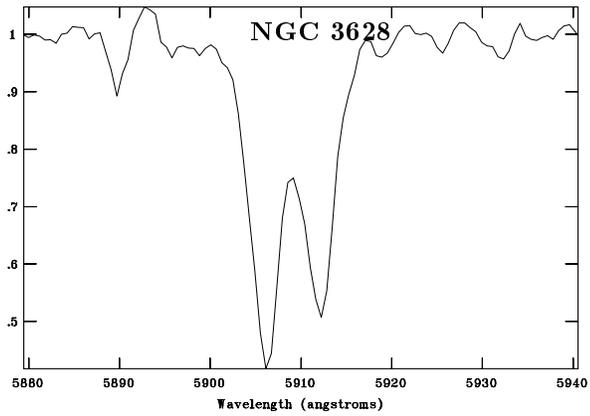
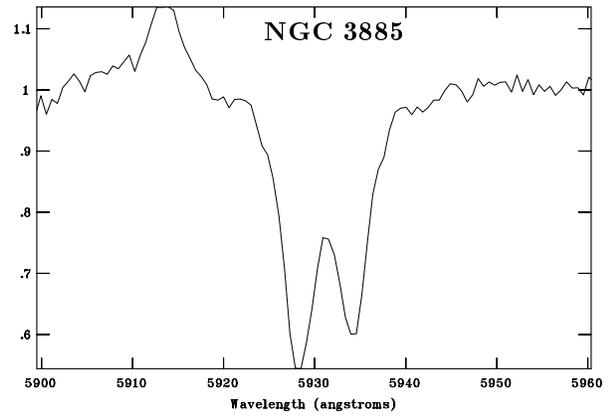
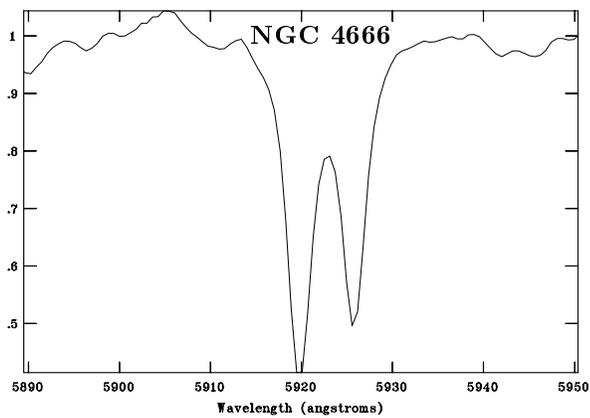
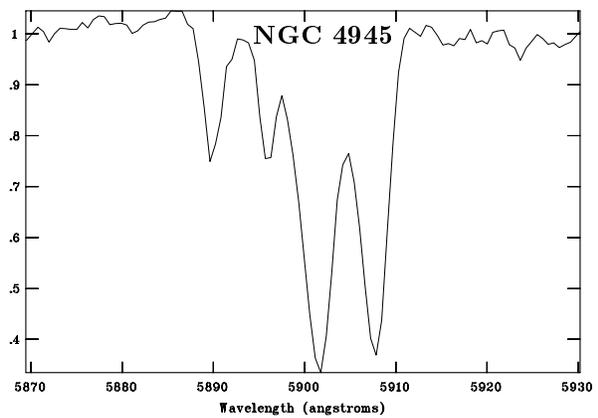
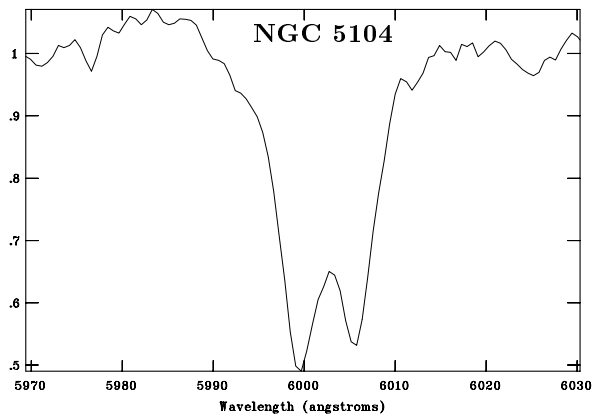
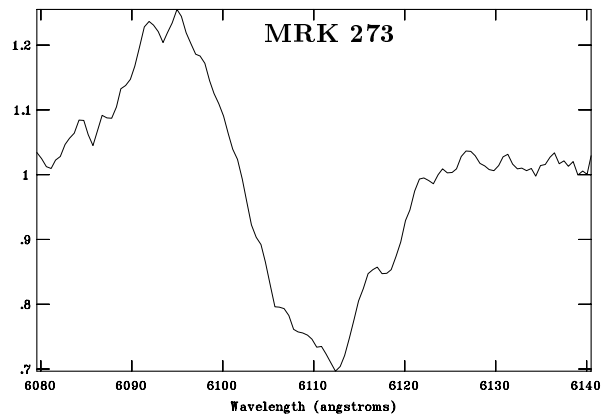
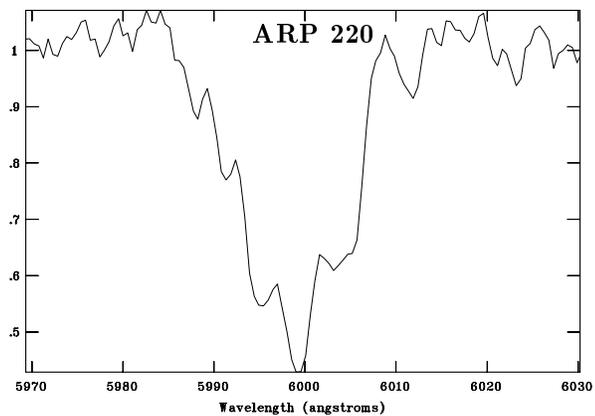
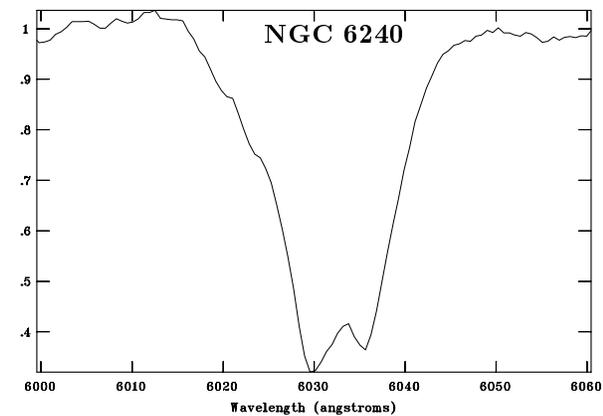
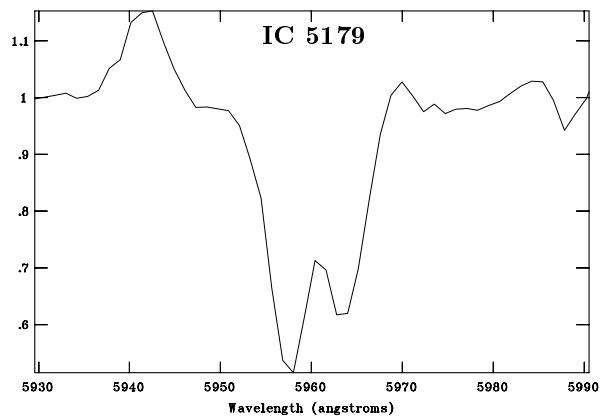
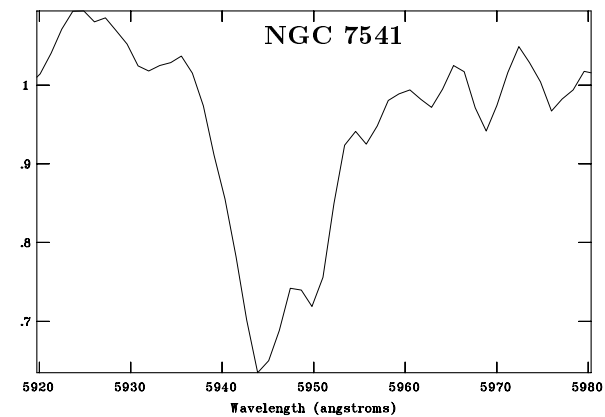
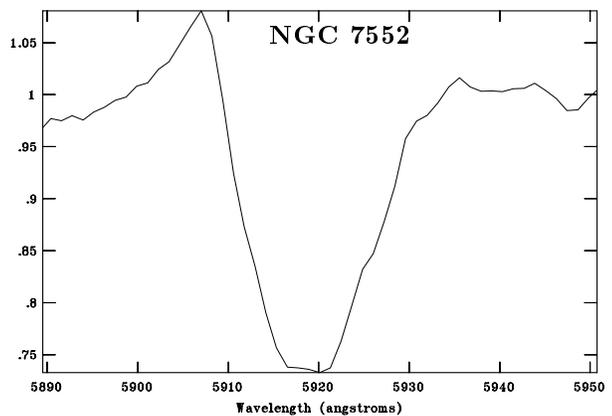
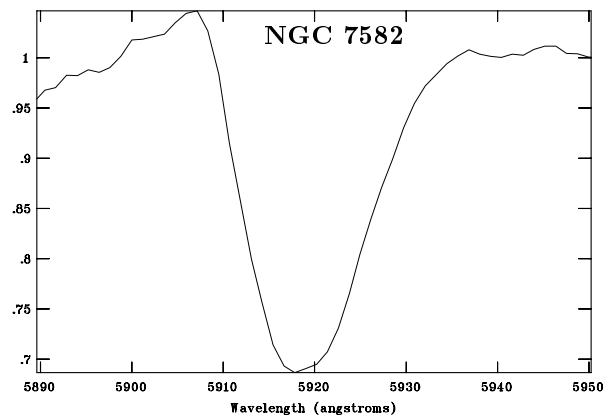
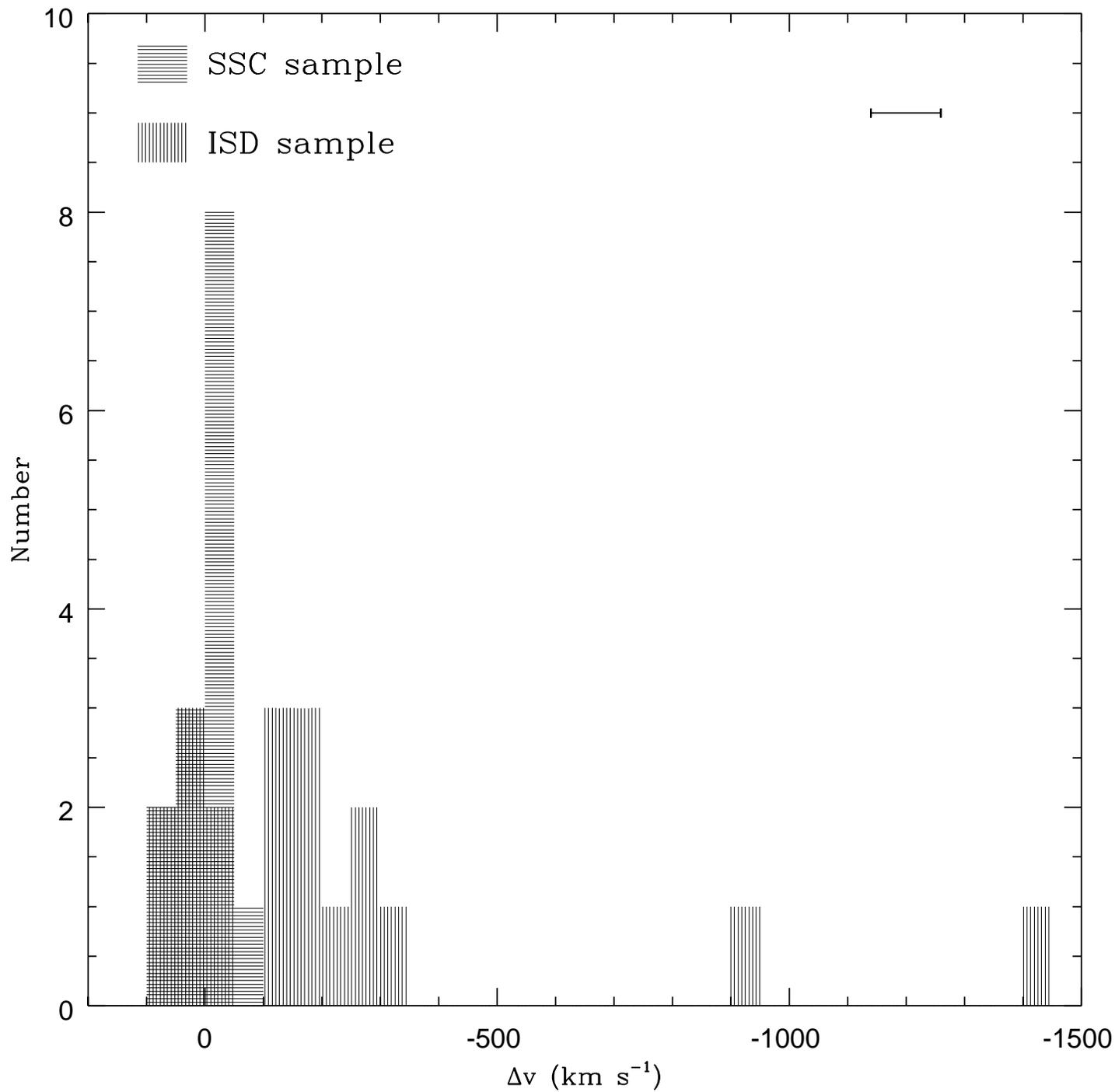
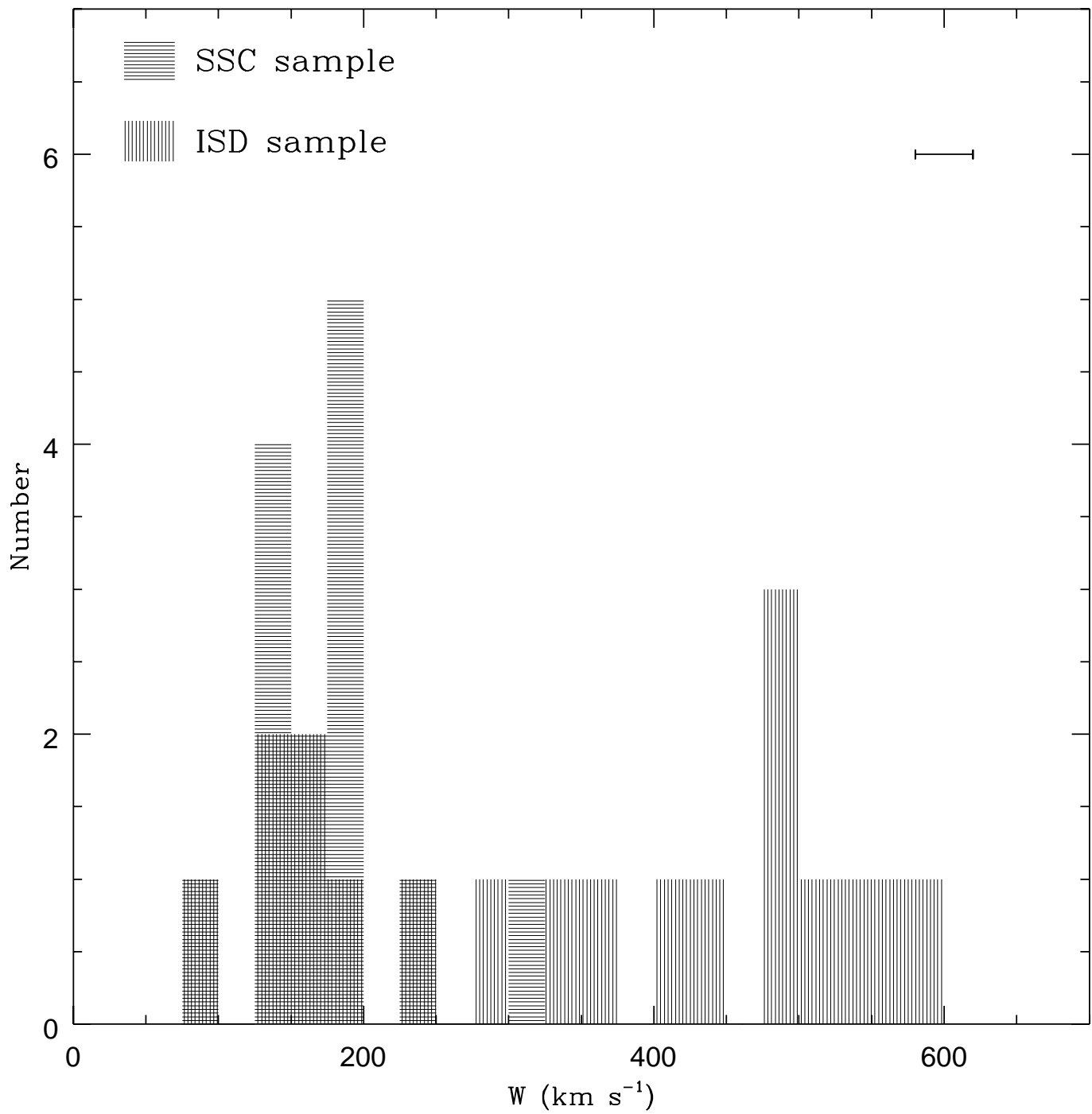
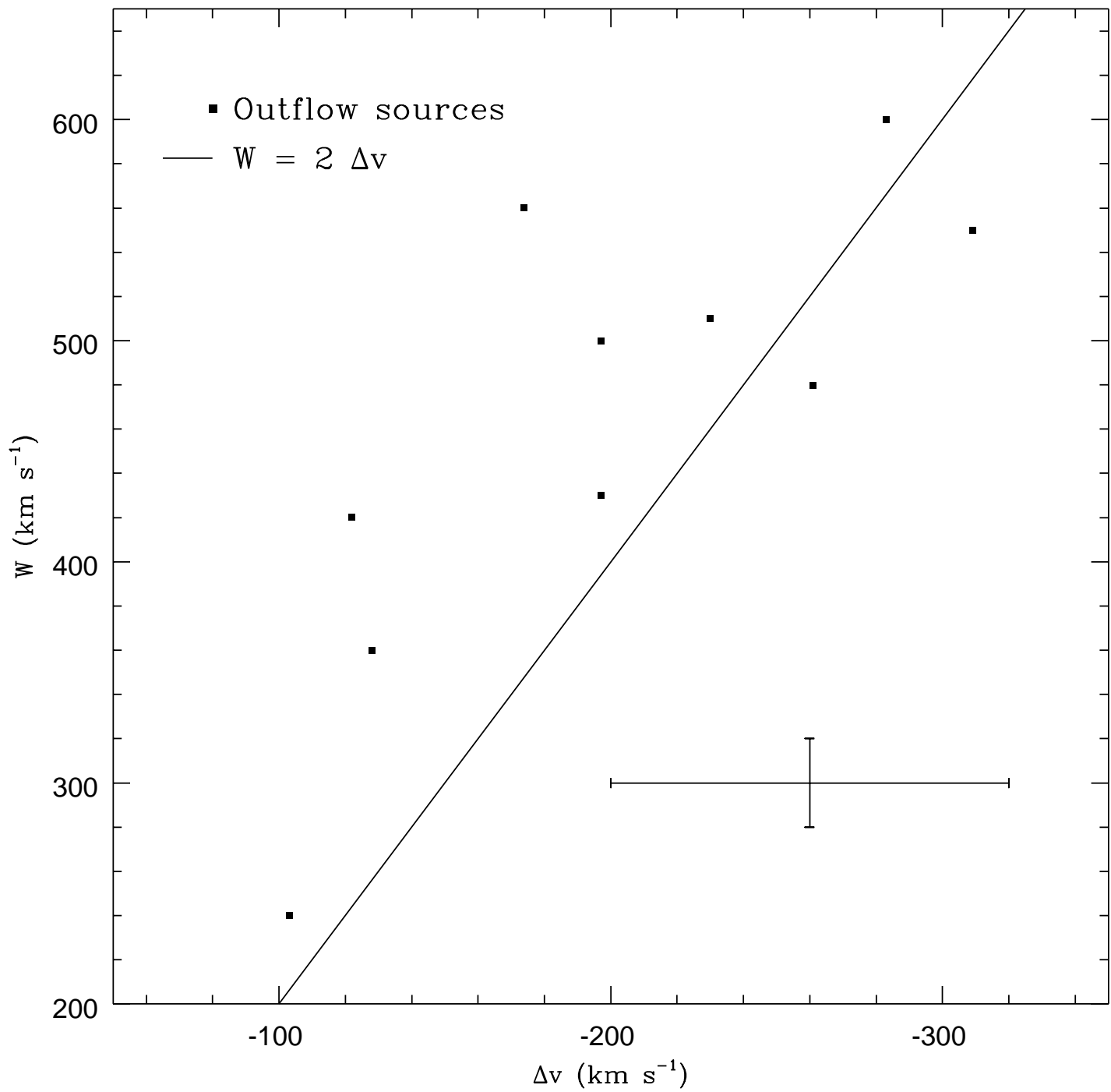
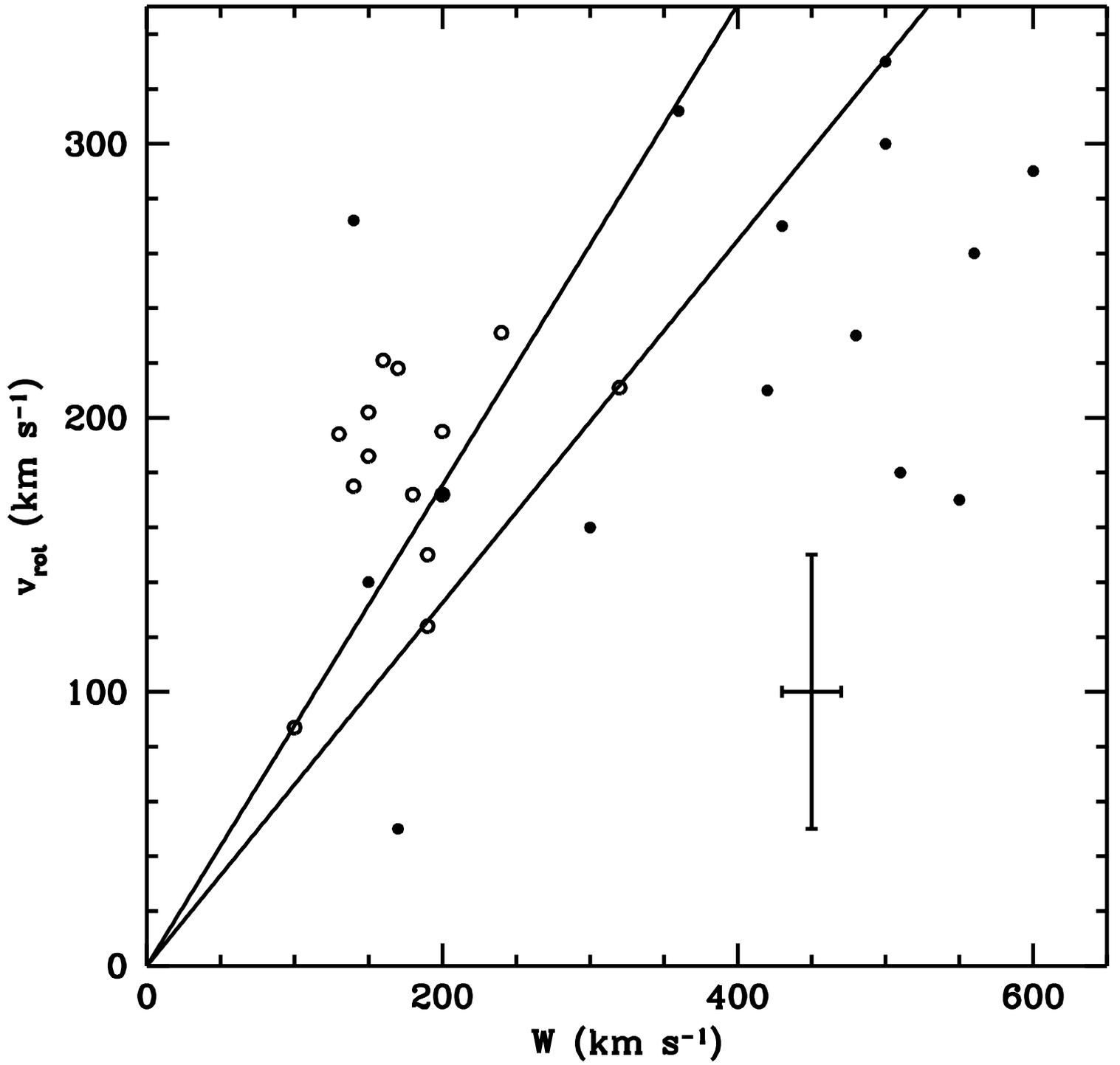
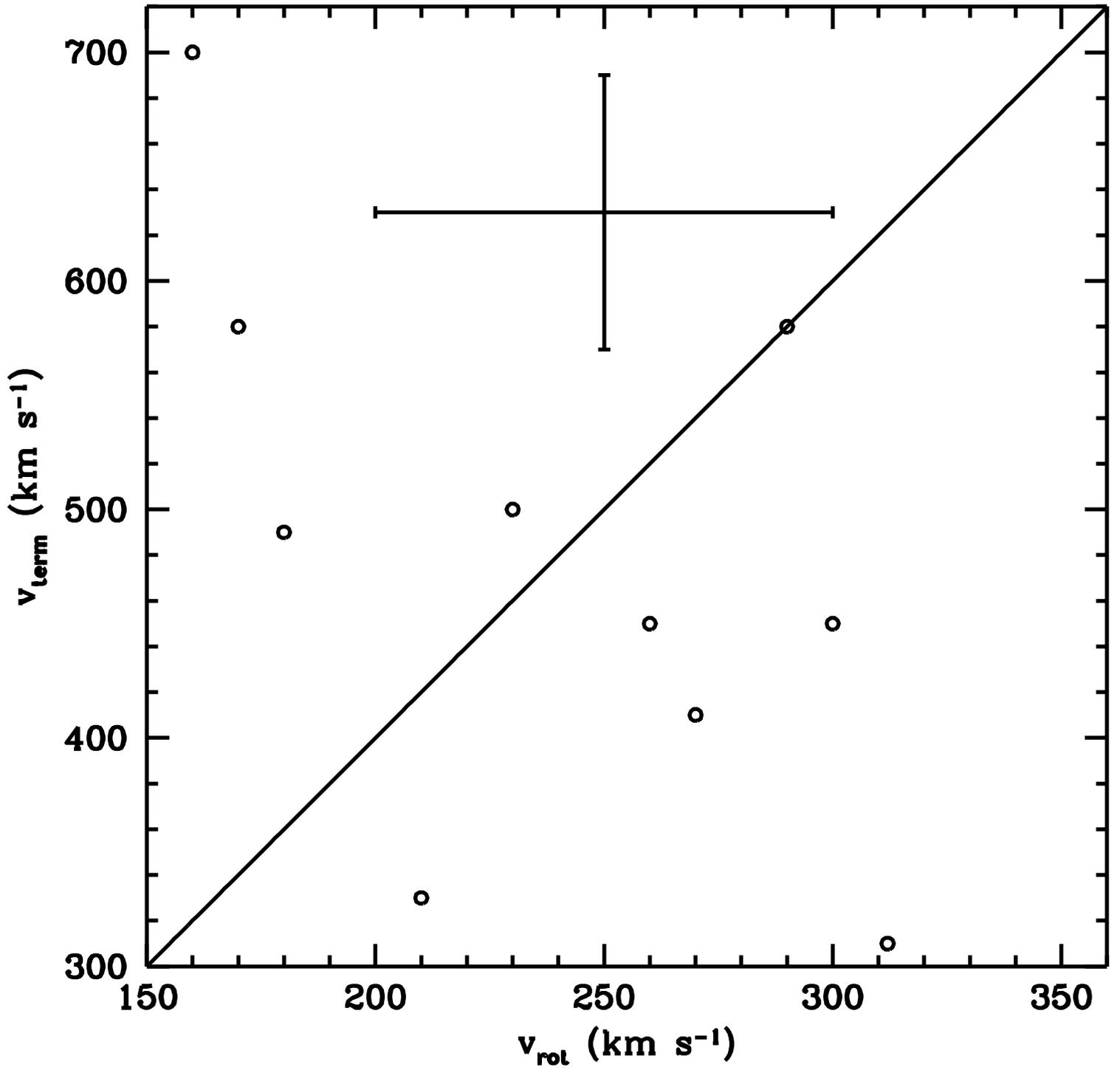
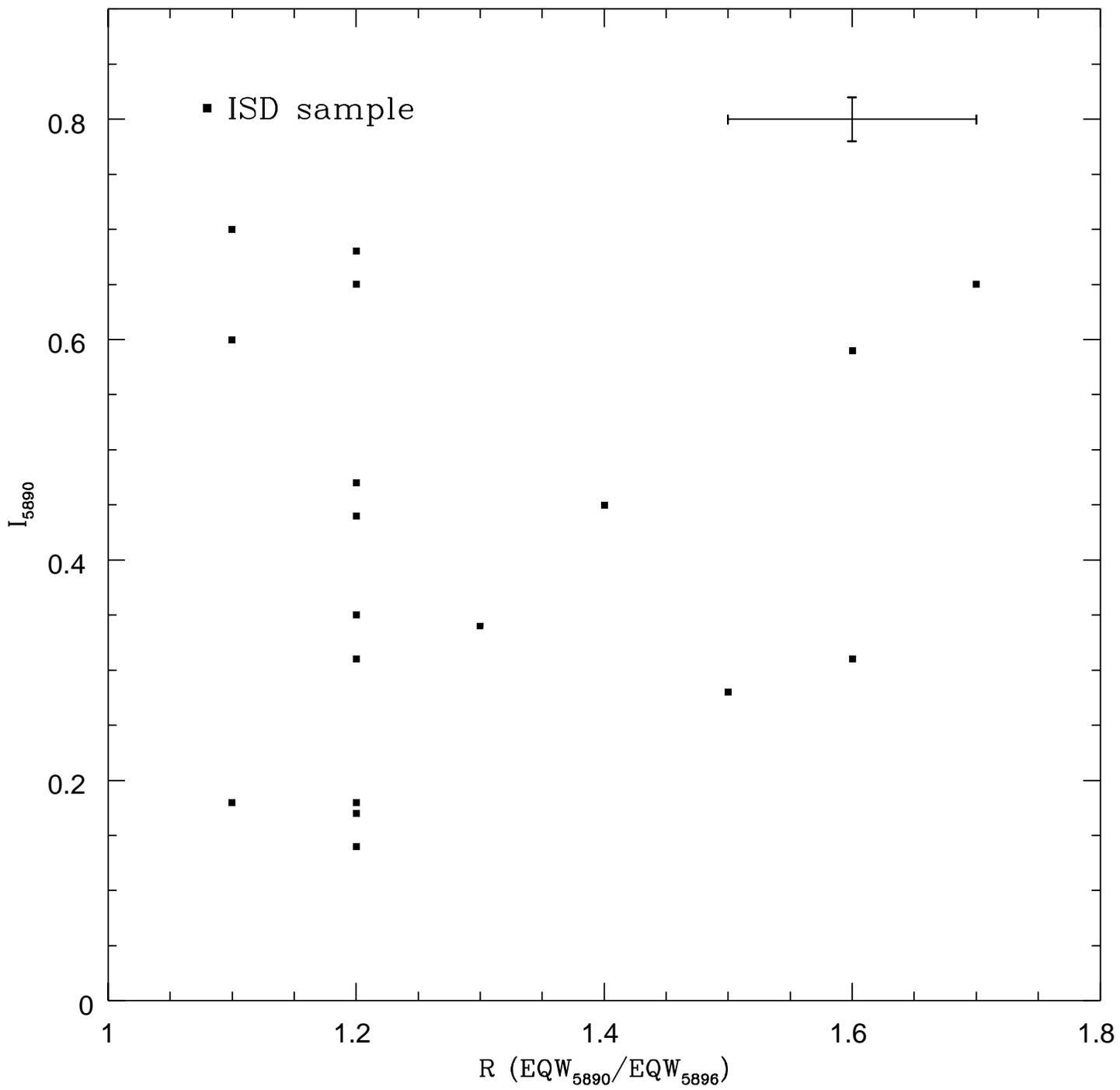
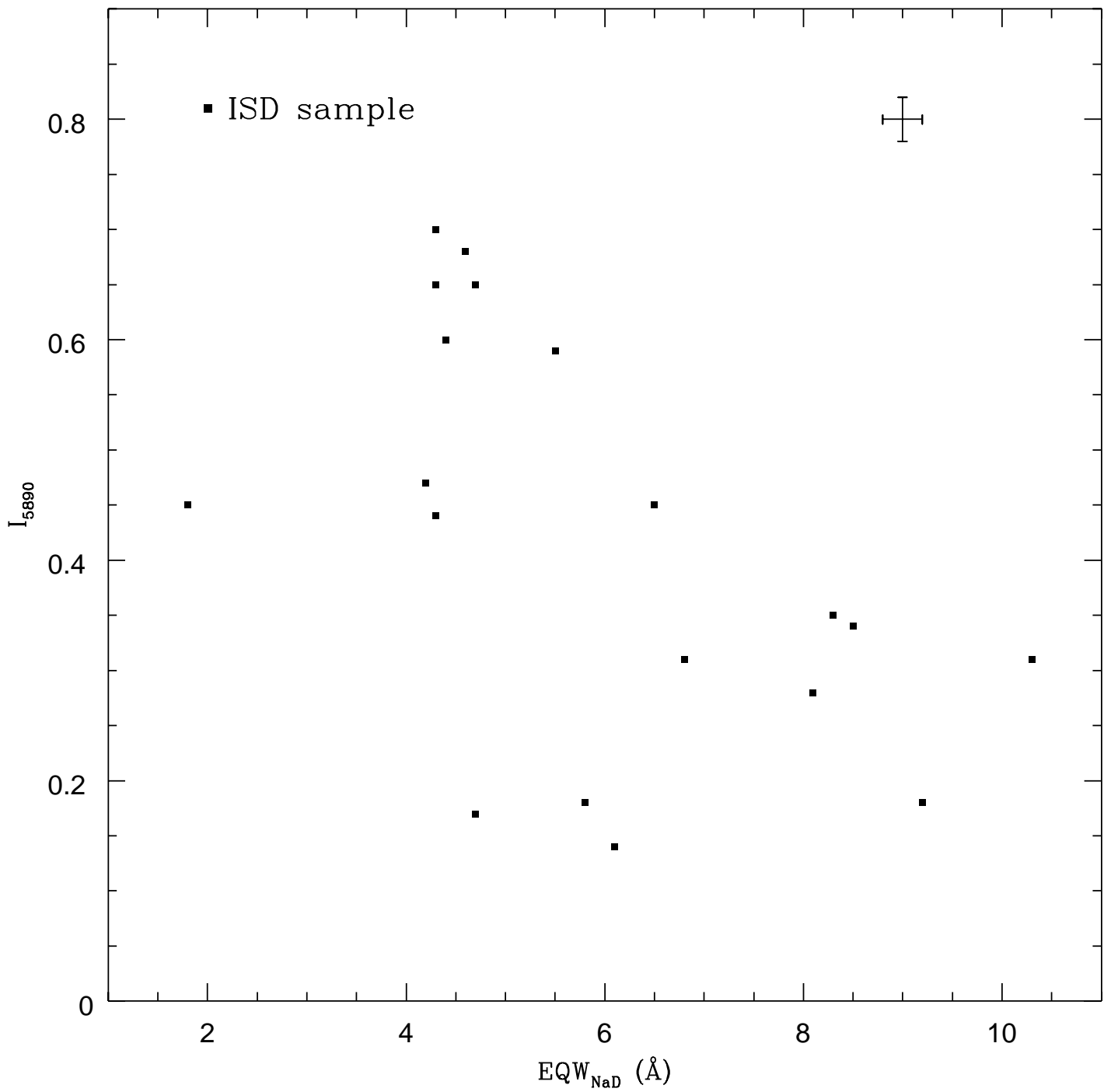
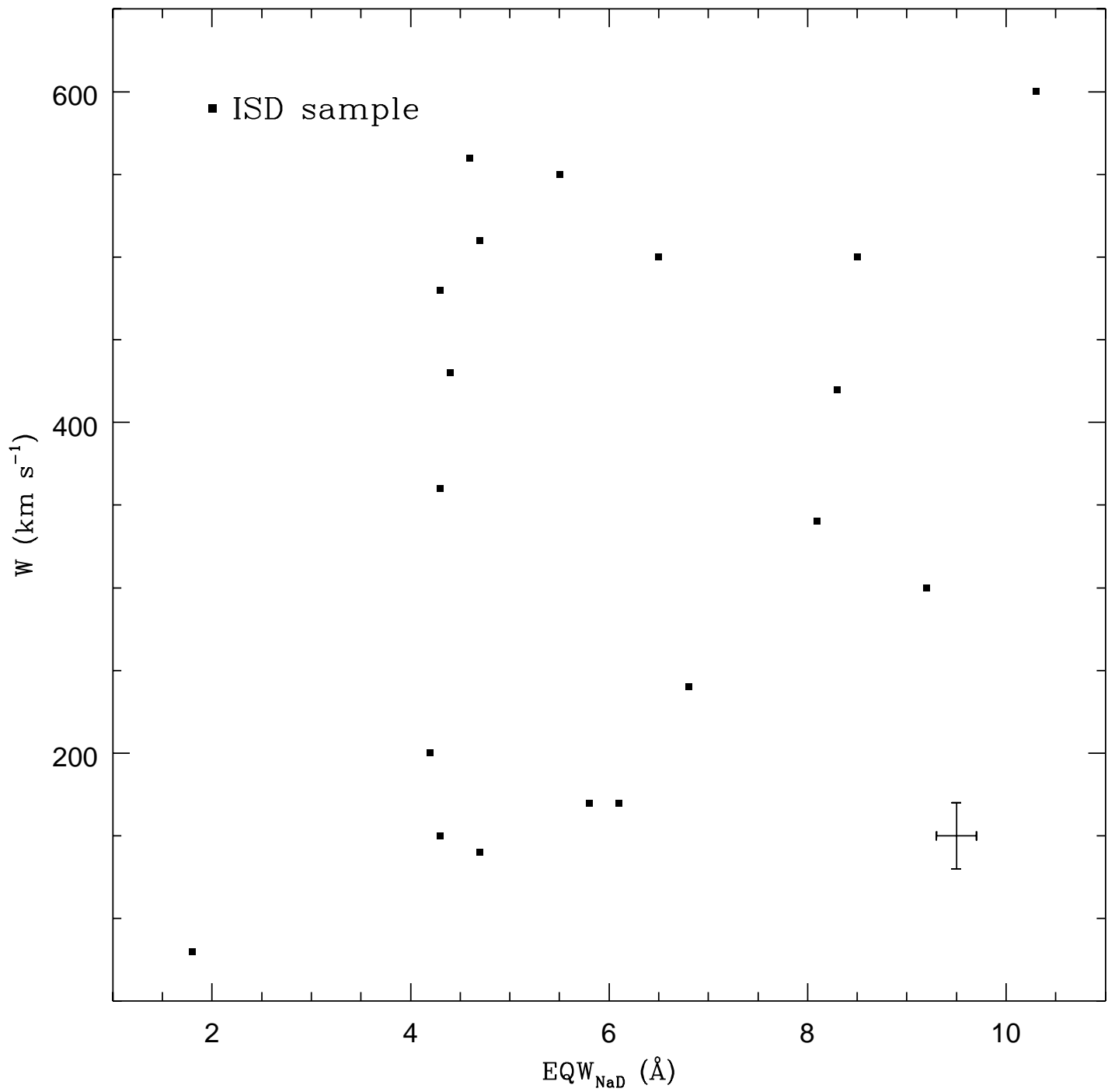
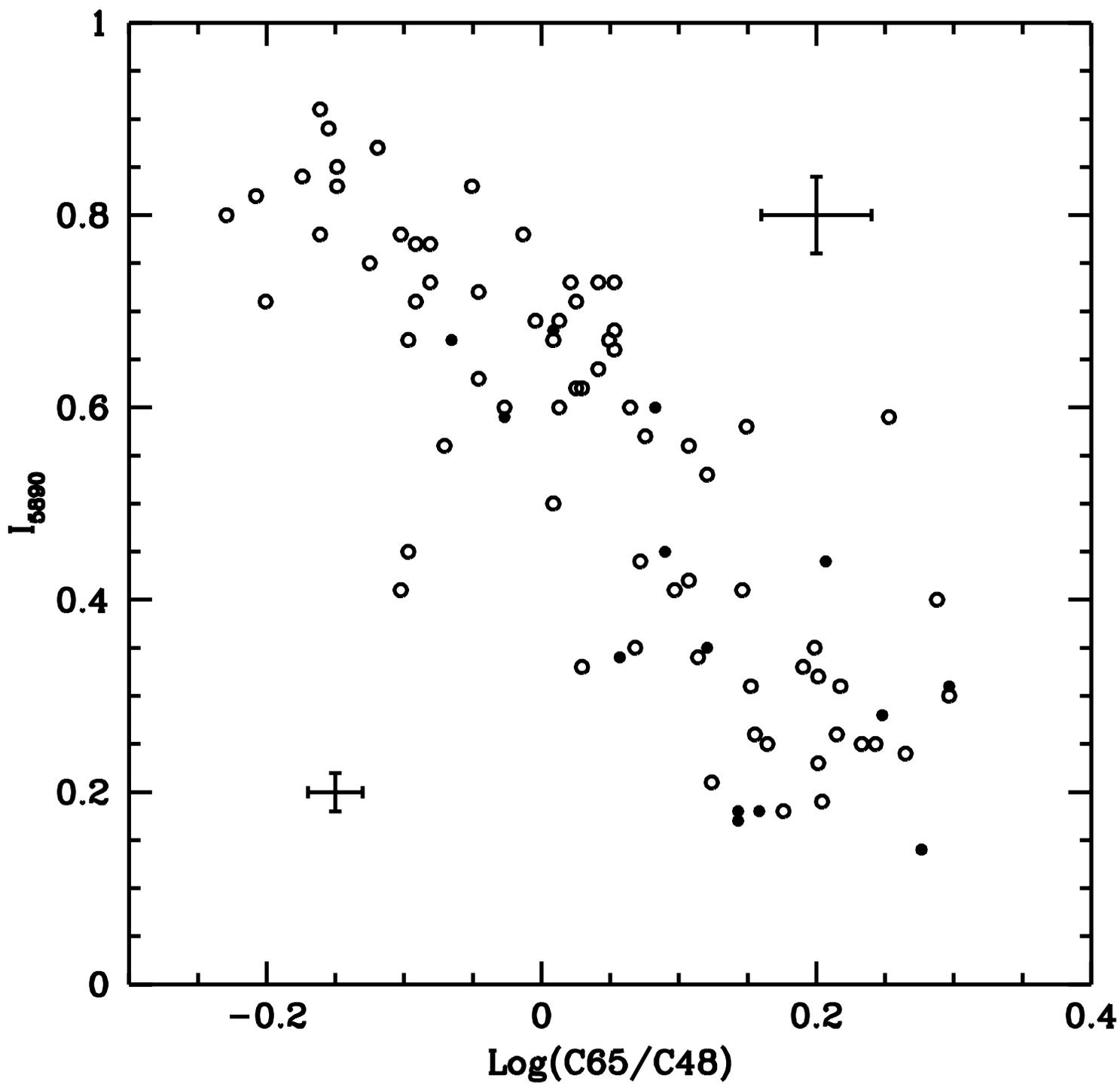
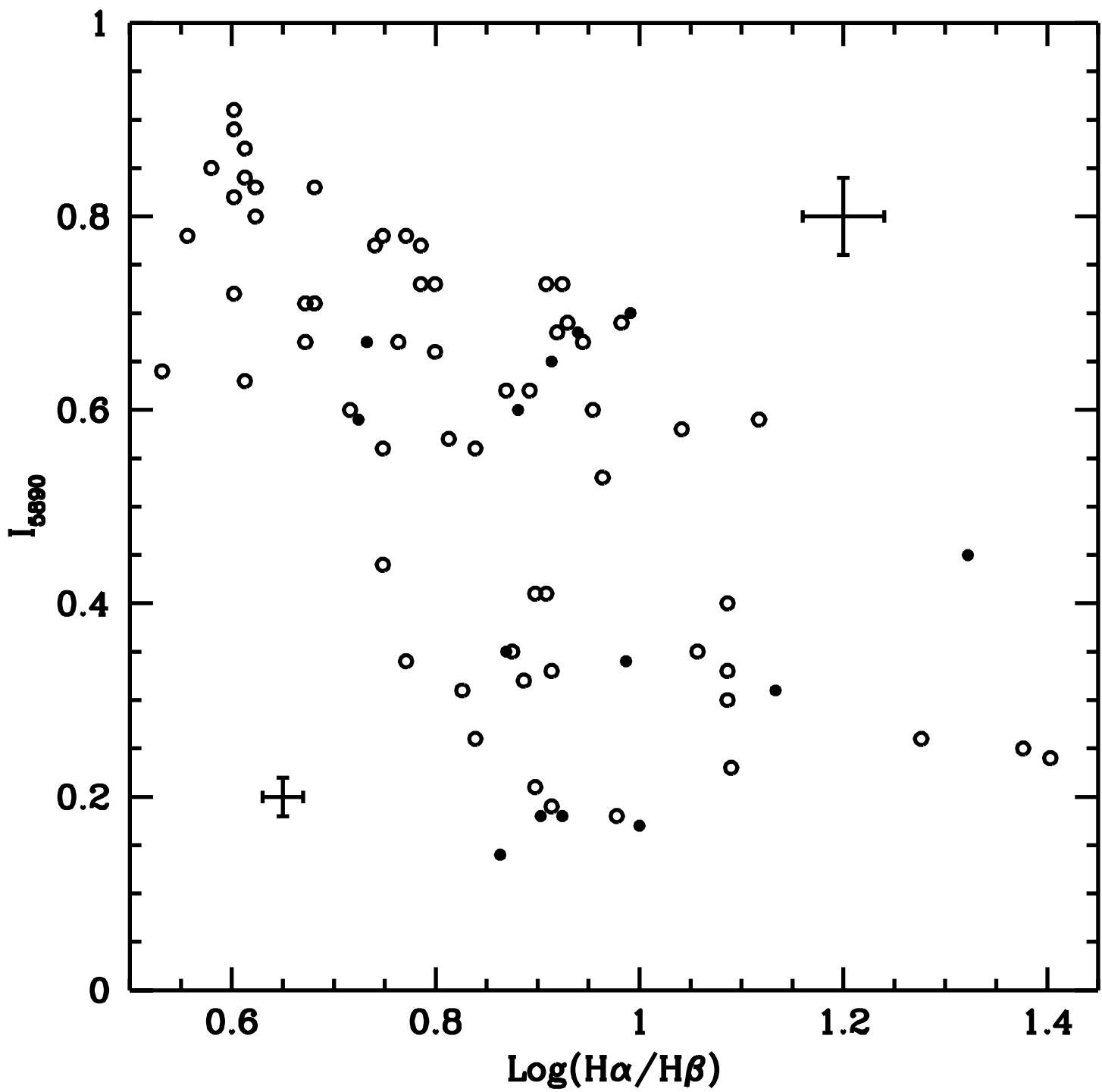
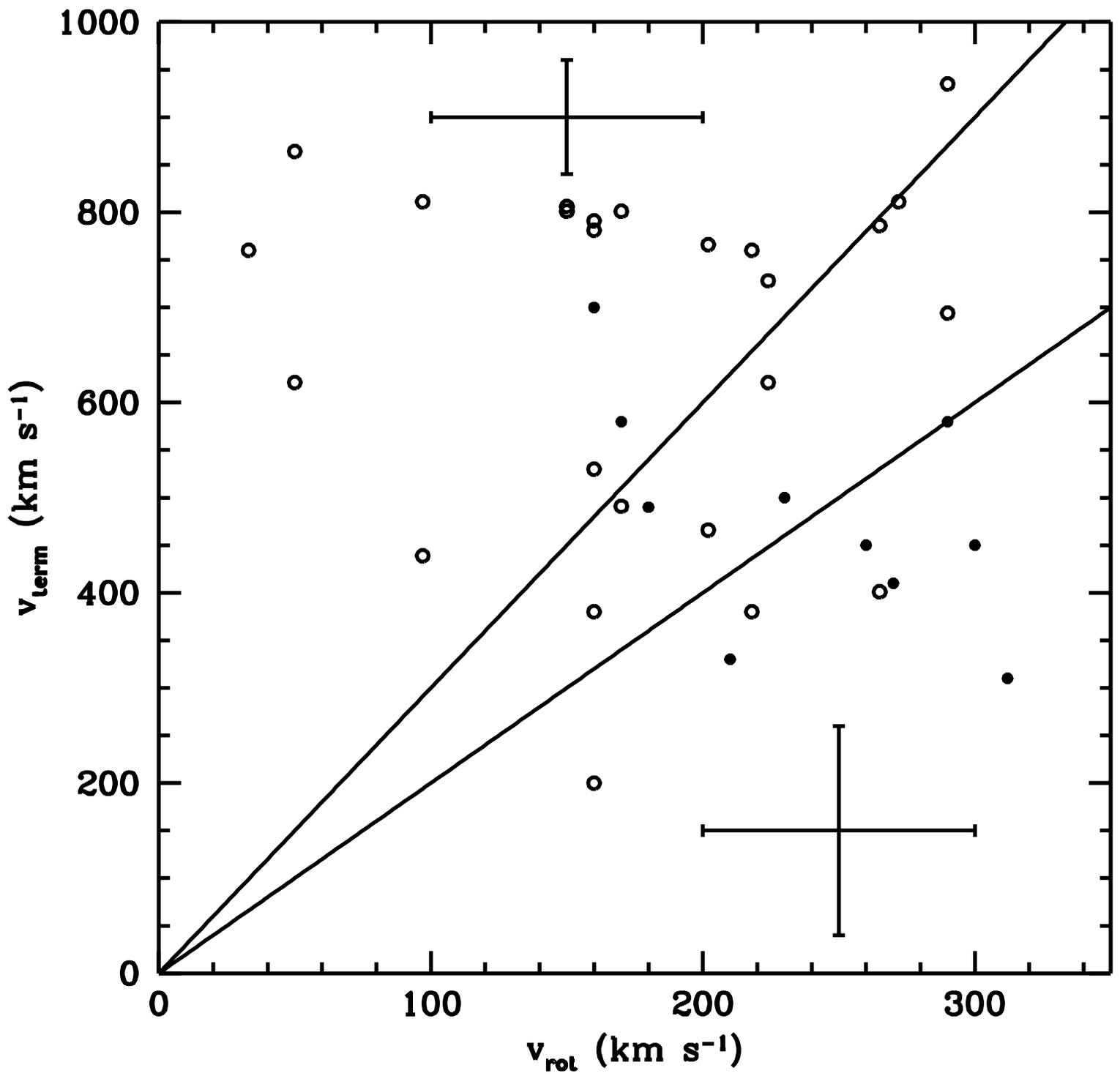
\end{document}

%% file: tab1.tex
\begin{deluxetable}{lrccccccc}
\tablecolumns{9}
\tablewidth{0pt}
\tablenum{1}
\tablecaption{Sample}
\tablehead{
\colhead{Galaxy}&\colhead{v$_{sys}$}&\colhead{}&
\colhead{log L$_{IR}$}&\colhead{M$_{B_T}$}&
\colhead{${a \over b}$}&\colhead{v$_{rot}$}&
\colhead{Sample}&\colhead{Run} \\
\colhead{(1)}&\colhead{(2)}&\colhead{}&
\colhead{(3)}&\colhead{(4)}&
\colhead{(5)}&\colhead{(6)}&
\colhead{(7)}&\colhead{(8)}}
\startdata
NGC 253       &   245&h,n &   10.5 &$-$19.8 & 4.1  &  202 h &   AHM,LH & 2 \nl
NGC 660       &   885&c   &   10.0 &$-$17.7 & 2.6  &  175 h &   AHM,LH & 5 \nl
IIIZw035      &  8294&e   &   11.5 &$-$19.5 & 2.0  &   87 h &   LH     & 1 \nl
IRAS02021--2104& 34630&e   &   12.0 &$-$21.1 & 2.5  &   ?    &   AHM    & 4 \nl
NGC 1134      &  3620&r   &   10.7 &$-$21.2 & 2.9  &  211 r &   LH     & 1 \nl
IRAS03514+1546&  6662&h,n &   11.1 &$-$20.8 & 1.1  &  270:h &   AHM    & 5,6 \nl
NGC1572       &  6142&r   &   11.2 &$-$21.4 & 2.0  &  312 r &   LH     & 1 \nl
NGC1614       &  4760&c   &   11.3 &$-$20.8 & 1.8  &  210:c &   AHM    & 4 \nl
IRAS04370--2416&  4537&r   &   11.0 &$-$20.0 & 2.5  &  172 r &   LH     & 1 \nl
NGC 1808      &  1001&r   &   10.6 &$-$20.1 & 1.7  &  160 r &   LH     & 3,4 \nl
IRAS05447--2114& 11977&e   &   11.0 &$-$18.7 & 2.0  &   ?    &   AHM,LH & 5 \nl
NGC 2146      &   916&r   &   10.7 &$-$19.3 & 1.8  &  272 h &   AHM,LH & 5 \nl
NGC 2966      &  2045&h,n &   10.2 &$-$19.2 & 2.6  &  124 r &   LH     & 4 \nl
M 82          &   214&c   &   10.5 &$-$18.5 & 2.6  &  50 h &   AHM.LH & 5,6 \nl
NGC 3094      &  2409&h,n &   10.4 &$-$19.5 & 1.4  &  150 h &   AHM    & 5 \nl
IRAS10173+0828& 14669&c   &   11.8 &$-$19.1 & 2.5  &  140 c &   AHM,LH & 5 \nl
NGC 3256      &  2801&c   &   11.5 &$-$21.3 & 1.8  &  170:c &   AHM    & 3,4 \nl
IRAS10502--1843& 16131&s   &   11.8 &$-$19.0 & 1.2  &   ?    &   AHM    & 4 \nl
IRAS10565+2448& 12923&c   &   12.0 &$-$20.7 & 1.3  &  300:c &   AHM    & 5,6 \nl
IRAS11119+3257& 56866&e   &   12.5 &  ?   & 1.2  &   ?    &   AHM    & 5 \nl
NGC 3628      &   843&h,n &   10.2 &$-$19.8 & 5.0  &  218 h &   LH     & 3 \nl
NGC 3885      &  1938&r   &   10.3 &$-$20.9 & 2.5  &  195 r &   LH     & 4 \nl
NGC 4666      &  1511&r   &   10.9 &$-$20.7 & 3.5  &  186 r &   LH     & 3 \nl
NGC 4945      &   560&h,n &   10.6 &$-$19.8 & 5.2  &  172 h &   LH     & 3 \nl
NGC 5104      &  5615&r   &   11.1 &$-$20.2 & 2.8  &  231 r &   LH     & 4  \nl
Mrk 273       & 11326&c,n &   12.2 &$-$21.0 & 2.0  &  260:c &   AHM    & 5,6 \nl
Arp 220       &  5441&c   &   12.2 &$-$20.7 & 1.3  &  330:c &   AHM    & 5 \nl
NGC 6240      &  7339&c,n &   11.7 &$-$21.6 & 1.9  &  290:c &   AHM,LH & 3 \nl
IC 5179       &  3424&r   &   11.0 &$-$20.9 & 2.1  &  194 r &   LH     & 1 \nl
NGC 7541      &  2714&r   &   10.8 &$-$20.3 & 2.8  &  221 r &   LH     & 1 \nl
NGC 7552      &  1585&n   &   10.8 &$-$20.2 & 1.2  &  230 h &   LH     & 1 \nl
NGC 7582      &  1575&n   &   10.6 &$-$20.2 & 2.4  &  180 r &   LH     & 1 \nl
\enddata
\end{deluxetable}

%% file: tab2.tex
\begin{deluxetable}{ccccccccc}
\tablecolumns{9}
\tablewidth{0pt}
\tablenum{2}
\tablecaption{Observing Runs}
\tablehead{
\colhead{Run Num}&\colhead{Date}&
\colhead{Obs}&\colhead{Tel}&
\colhead{Spec}&\colhead{Detector}&
\colhead{pixels}&\colhead{Slit}&
\colhead{Res} \\
\colhead{(1)}&\colhead{(2)}&
\colhead{(3)}&\colhead{(4)}&
\colhead{(5)}&\colhead{(6)}&
\colhead{(7)}&\colhead{(8)}&
\colhead{(9)}}
\startdata
1  & 11/90& LCO &2.5m& ModSpec       &TEK1024  &0.68$\times$1.2  &2  &3.4 \nl
2  & 10/92& CTIO& 4m&Blue Air Schmidt&Reticon1 &0.77$\times$0.93 &2  &2.1 \nl
3  & 3/93 & CTIO& 4m&Folded Schmidt  &TEK1024  &0.79$\times$0.60 &2.2&1.8 \nl
   &      &     &   &                &         &0.79$\times$1.98 &2.2&5.9 \nl
4  & 1/94 & CTIO& 4m&Folded Schmidt  &TEK1024  &0.82$\times$0.6  &2  &1.8 \nl
5  & 1/94 & KPNO& 4m&RC Spec         &T2KB     &0.69$\times$0.5  &2  &1.1 \nl
6  & 1/88 & KPNO& 4m&RC Spec         &TI2      &0.90$\times$3.4  &2 &13.5 \nl
\enddata
\tablecomments{
Col. (7) --- Pixel size in arcsec by \AA.
Col. (8) --- Slit width in arcsec.
Col. (9) --- Spectral resolution (FWHM) in \AA\ for $NaD$.}
\end{deluxetable}

%% file: tab3.tex
\begin{deluxetable}{lrccccccc}
\tablecolumns{9}
\tablewidth{0pt}
\tablenum{3}
\tablecaption{Measured Properties}
\tablehead{
\colhead{Galaxy}&\colhead{f$_{\star}$}&
\colhead{v$_{NaD}$}&\colhead{$\Delta$v}&
\colhead{W}&\colhead{EQW}&
\colhead{I$_{5890}$}&\colhead{R}&
\colhead{W$_{H\alpha}$} \\
\colhead{(1)}&\colhead{(2)}&
\colhead{(3)}&\colhead{(4)}&
\colhead{(5)}&\colhead{(6)}&
\colhead{(7)}&\colhead{(8)}&
\colhead{(9)}}
\startdata
NGC253         &  50\%  &  193&  $-$52&   150&6.0 & 0.04  &1.1  &280 \nl
NGC660         &  50\%  &  878&   $-$7&   140&4.6 & 0.22  &1.2  &190 \nl
IIZw035        &  50\%  & 8365&     69&   100&2.5 & 0.37  &1.4  &310 \nl
IRAS02021--2104&$<$20\% &34692&     56&   340& 8.1&  0.28&  1.5&  420 \nl
NGC1134        &  70\%  & 3620&      0&  320 &4.4  &0.57  &1.2  &220 \nl
IRAS03514+1546 &  30\%  & 6461& $-$197&  430 &4.4  &0.60  &1.1  &170 \nl
NGC1572        &  30\%  & 6011& $-$128&  360 &4.3  &0.65  &1.2  &330 \nl
NGC1614        &$<$10\% & 4636& $-$122&   420& 8.3&  0.35&  1.2&  300 \nl
IRAS04370--2416&  40\%  & 4525&  $-$12&  200 &2.4  &0.67  &1.3  &170 \nl
NGC1808        &  20\%  & 1013&     12&  300 &9.2  &0.18  &1.1  &260 \nl
IRAS05447--2114&$<$30\% &12072&     91&   200& 4.2&  0.47&  1.2&  290 \nl
NGC2146        &  30\%  &  930&     14&  140 &4.7  &0.17  &1.2  &120 \nl
NGC2966        &  50\%  & 2043&   $-$2&  190 &3.8  &0.45  &1.2  &210 \nl
M 82           &$<$20\% &  204&  $-$10&   170& 5.8&  0.18&  1.2&  100 \nl
NGC3094        &  50\%  & 2394&  $-$15&  190 &3.9  &0.48  &1.3  &120 \nl
IRAS10173+0828 &$<$30\% &14708&     37&   150& 4.3&  0.44&  1.2&  200 \nl
NGC3256        &  20\%  & 2489& $-$309&   550&5.5  &0.59  &1.6  &210 \nl
IRAS10502--1843& $<$20\%&16022& $-$103&   240& 6.8&  0.31&  1.2&  310 \nl
IRAS10565+2448 &$<$20\% &12717& $-$197&   500& 8.5&  0.34&  1.3&  210 \nl
IRAS11119+3257 &$<$10\% &55755& $-$934&   170& 6.1&  0.14&  1.2& 1500 \nl
               &        &55189&$-$1410&    80& 1.8&  0.45&  ...&      \nl
NGC3628        &  70\%  &  812&  $-$31&   170& 4.5 & 0.29 & 1.2 & 120 \nl
NGC3885        &  70\%  & 1962&     24&   200& 4.1 & 0.46 & 1.2 & 300 \nl
NGC4666        &  70\%  & 1521&     10&   150& 4.3 & 0.25 & 1.2 & 200 \nl
NGC4945        &  50\%  &  622&     62&   180& 5.5 & 0.33 & 1.1 & 390 \nl
NGC5104        &  50\%  & 5603&  $-$12&   240& 5.3 & 0.41 & 1.1 & 470 \nl
Mrk273         &  30\%  &11145& $-$174&   560& 4.6 & 0.68 & 1.2 & 520 \nl
Arp 220        &  30\%  & 5422&  $-$19&   500& 6.5 & 0.45 & 1.4 & 610 \nl
NGC6240        &  20\%  & 7049& $-$283&   600&10.3 & 0.31 & 1.6 & 890 \nl
IC5179         &  40\%  & 3431&      7&   130& 4.5 & 0.28 & 1.2 & 180 \nl
NGC7541        &  50\%  & 2695&  $-$19&   160& 3.8 & 0.42 & 1.2 & 260 \nl
NGC7552        &  30\%  & 1323& $-$261&   480& 4.3 & 0.70 & 1.1 & 140 \nl
NGC7582        &  30\%  & 1344& $-$230&   510& 4.7 & 0.65 & 1.7 & 190 \nl
\enddata
\end{deluxetable}

%% file: tab4.tex
\begin{deluxetable}{lccccccc}
\tablecolumns{6}
\tablewidth{0pt}
\tablenum{4}
\tablecaption{Miscellaneous Properties of ``ISD'' Sub-Sample}
\tablehead{
\colhead{Galaxy}&\colhead{${H\alpha \over H\beta}$}&
\colhead{${C65 \over C48}$}&\colhead{Size}&
\colhead{v$_{term}$}&\colhead{v$_{rot}$} \\
\colhead{(1)}&\colhead{(2)}&
\colhead{(3)}&\colhead{(4)}&
\colhead{(5)}&\colhead{(6)}}
\startdata
IRAS02021--2104& ...     &    1.77     &     ...      &   ...    &     ... \nl
IRAS03514+1546&7.6    &    1.21     &    4.4     &  410   &    270:  \nl
NGC1572       & ...     &      ...      &   2.4x3.2  &  310   &    312 \nl
NGC1614       &7.4    &    1.32     &  3.2x3.2   &  330   &    210: \nl
IRAS04370--2416&5.4    &    0.86     &     ...      &   ...    &    172 \nl
NGC 1808      &8.0    &    1.39     &    3.7     &  700*  &    160 \nl
IRAS05447--2114& ...&       ...     &      ...     &    ...   &      ...  \nl
NGC 2146      &10.0&    1.39     &     ...      &   ...    &    272 \nl
M 82          &8.4&    1.44     &     ...      &   ...    &    50 \nl
IRAS10173+0828&...&    1.61     &     ...      &   ...    &    140 \nl
NGC 3256      &5.3&    0.94     &   5.6x1.8  &  580   &    170: \nl
IRAS10502--1843&...&      ...      &    7.7     &  220   &    ... \nl
IRAS10565+2448&9.7&    1.14     &    8.5     &  450   &    300: \nl
IRAS11119+3257&7.3&    1.89     &    $<$6      & 1450   &     ... \nl
Mrk 273       &8.7&    1.02     &     3      &  450   &    260: \nl
Arp 220       &21 &    1.23     &     ...      &   ...    &    330: \nl
NGC 6240      &13.6&    1.98     &     9      &  580   &    290: \nl
NGC 7552      &9.8&     ...       &   1.2x1.2  &  500   &    230 \nl
NGC 7582      &8.2&     ...       &   1.0x1.0  &  490   &    180 \nl
\enddata
\end{deluxetable}

%% file: ms.bbl
\begin{thebibliography}{}

\bibitem[]{}Aalto, S., Black, J., Johansson, L., \& Booth, R. 1991, A\&A, 249,
323
\bibitem[]{}Aguirre, A. 1999a, ApJL, 512, L19
\bibitem[]{}Aguirre, A. 1999b, ApJ, 525, 583
\bibitem[]{}Aguirre, A., \& Haiman, Z. 2000, astro-ph/9907039
\bibitem[]{}Alton, P.,  Davies, J., \& Bianchi, S. 1999, A\&A, 343, 51
\bibitem[]{}Alton, P., Bianchi, S., Rand, R.,  Xilouris, E., Davies, J., \&
Trewhella, M. 1998, ApJL, 507, L125
\bibitem[]{}Alton, P., Draper, P., Gledhill, T., Stockdale, D., Scarrott, S.,
\& Wolstencroft, R. 1994, MNRAS, 270, 238
\bibitem[]{}Antonucci, R. 1993, ARA\&A, 31, 473
\bibitem[]{}Armus, L., Heckman, T., \& Miley, G. 1987, AJ, 94, 831
\bibitem[]{}Armus, L., Heckman, T., \& Miley, G. 1989, ApJ, 347, 727
\bibitem[]{}Armus, L., Heckman, T., \& Miley, G. 1990, ApJ, 364, 471
\bibitem[]{}Awaki, M., Ueno, S., Koyama, K., Tsuru, T., \& Iwasawa, K. 1996,
PASJ, 48, 409
\bibitem[]{}Barger, A., Cowie, L., Smail, I., Ivison, R., Blain, A., \&
Kneib, J.-P. 1999, AJ, 117, 2656
\bibitem[]{}Barlow, T., \& Sargent, W. 1997, AJ, 113, 136
\bibitem[]{}Bender, R., Burstein, D., \& Faber, S. 1993, ApJ, 411, 153
\bibitem[]{}Bica, E., Pastoriza, M., Da Silva, L., Dottori, H., \& Maia, M.
1991, AJ, 102, 1702
\bibitem[]{}Binney, J. \& Tremaine, S. 1987, ``Galactic Dynamics'' (Princeton
University Press, Princeton, NJ)
\bibitem[]{}Bruzual, A.G. \& Charlot, S.1993, ApJ, 405, 538
\bibitem[]{}Bland-Hawthorn, J. 1995, PASA, 12, 190
\bibitem[]{}Calzetti, D. \& Heckman, T. 1999, ApJ, 519, 27
\bibitem[]{}Calzetti, D. 1997, AJ, 113, 162
\bibitem[]{}Carollo, M. \& Danziger, I.J. 1994, MNRAS, 270, 523
\bibitem[]{}Casoli, F., Dupraz, C., Combes, F., \& Kazes, I. 1991, A\&A, 251, 1
\bibitem[]{}Cen, R., \& Ostriker, J. 1999, ApJ, 514, 1
\bibitem[]{}Chevalier, R., \& Clegg, A. 1985, Nature, 317, 44
\bibitem[]{}Crenshaw, D.M., Kraemer, S., Boggess, A., Maran, S., Mushotzky, R.,
Wu, C.-C. 1999, ApJ, 516, 750
\bibitem[]{}Dahari, O. \& DeRobertis, M. 1988, ApJS, 67, 249
\bibitem[]{}Dahlem, M. 1997, PASP, 109, 1298
\bibitem[]{}Dahlem, M., Weaver, K., \& Heckman, T. 1998, ApJS, 118, 401
\bibitem[]{}Davies, J., Phillips, S., Trewhella, M., \& Alton, P. 1997,
MNRAS, 291, 59
\bibitem[]{}Della Ceca, R., Griffiths, R., Heckman, T., Lehnert, M., \&
Weaver, K. 1999, ApJ, 514, 772
\bibitem[]{}Della Ceca, R., Griffiths, R., \& Heckman, T. 1997, ApJ, 485, 581
\bibitem[]{}Della Ceca, R., Griffiths, R., Heckman, T., \& MacKenty, J. 1996,
ApJ, 469, 662
\bibitem[]{}Della Ceca, R., Griffiths, R., Heckman, T., Lehnert, M., \&
Weaver, K. 1999, ApJ, 514, 772
\bibitem[]{}Downes, D. \& Solomon, P. 1998, ApJ, 507, 615
\bibitem[]{}Draper, P. Done, C., Scarrott, S., \& Stockdale, D. 1995, MNRAS,
277, 1430
\bibitem[]{}Elfhag, T., Booth, R., Hoeglund, B., Johansson, L., \& Sandqvist,
A. 1996, A\&AS, 115, 439
\bibitem[]{}Elmegreen, B. 1999, ApJ, 517, 103
\bibitem[]{}Ferrera, A., Ferrini, F., Barsella, B., \& Franco, J. 1991, ApJ,
381, 137
\bibitem[]{}Ferrara, A. 1998, in ``Proc. IAU Symp. 166'', Ed. D. Breitschwerdt,
M. Freyberg, \& J. Truemper (Springer-Verlag: Berlin), p. 371
\bibitem[]{}Ferrera, A., Nath, B., Sethi, S., \& Shchekinov, Y. 1999, MNRAS,
303, 301
\bibitem[]{}Franx, M. 1993, in Proc. IAU Symp. 153, Ed. H. Dejonghe \& H.
Habing (Kluwer: Dordrecht), p. 243
\bibitem[]{}Franx, M. \& Illingworth, G. 1990, ApJL, 359, L41
\bibitem[]{}Franx, M., Illingworth, G., Kelson, D., van Dokkum, P., \& Tran,
K.-V. 1997, ApJL, 486, L75
\bibitem[]{}Fukugita, M., Hogan, C., \& Peebles, J. 1998, ApJ, 503, 518
\bibitem[]{}Gibson, B., Loewenstein, M., \& Mushotzky, R. 1997, MNRAS, 290, 623
\bibitem[]{}Giroux, M., \& Shull, S.M. 1997, AJ, 113, 1505
\bibitem[]{}Gonzalez-Delgado, R., Leitherer, C., Heckman, T., Lowenthal, J.,
Ferguson, H., \& Robert, C. 1998a, ApJ, 495, 698
\bibitem[]{}Gonzalez-Delgado, R., Heckman, T., Leitherer, C., Meurer, G.,
Krolik, J., Wilson, A., Kinney, A., \& Koratkar, A. 1998b, ApJ, 505, 174
\bibitem[]{}Gonzalez-Delgado, R., Heckman, T., \& Leitherer, C. 2000, in
preparation.
\bibitem[]{}Gordon, K., Calzetti, D., \& Witt, A. 1997, ApJ, 487, 625
Habing (Kluwer: Dordrecht), p. 243
\bibitem[]{}Hamann, F., Barlow, T., Junkkarinen, V., \& Burbidge, E.M. 1997,
ApJ, 478, 80
\bibitem[]{}Hartquist, T., Dyson, J., \& Williams, R. 1997, ApJ, 482, 182
\bibitem[]{}Heckman, T. 1980, A\&A, 87, 142
\bibitem[]{}Heckman, T., \& Leitherer, C. 1997, AJ, 114, 69
\bibitem[]{}Heckman, T. M., Armus, L., \& Miley, G. K. 1990, ApJS, 74, 833
\bibitem[]{}Heckman, T., Lehnert, M., \& Armus, L. 1993, in ``The Environment
and Evolution of Galaxies'', ed. J.M. Shull \& H. Thronson (Kluwer: Dordrecht),
p. 455
\bibitem[]{}Heckman, T., Balick, B., van Breugel, W., \& Miley, G. 1983, AJ,
88, 583
\bibitem[]{}Heckman, T., Krolik, J., Meurer, G., Calzetti, D., Koratkar, A.,
Leitherer, C., Robert, C., \& Wilson, A. 1995, ApJ, 452, 549
\bibitem[]{}Heckman, T., Gonzalez-Delgado, R., Leitherer, C., Meurer, G.,
Krolik, J. Wilson, A., Koratkar, A., \& Kinney, A. 1997, ApJ, 482, 114
\bibitem[]{}Heckman, T., Robert, C., Leitherer, C., Garnett, D., \& van der
Rydt, F. 1998, ApJ, 503, 646
\bibitem[]{}Heckman, T., Armus, L., Weaver, K., \& Wang, J. 1999, ApJ, 517, 130
\bibitem[]{}Heckman, T., \& Lehnert, M. 2000, submitted to ApJ
\bibitem[]{}Heisler, J. \& Ostriker, J. 1988, ApJ, 332, 543
\bibitem[]{}Herbig, G. 1993, ApJ, 407, 142
\bibitem[]{}Howk, J.C., \& Savage, B. 1997, AJ, 114, 2463
\bibitem[]{}Howk, J.C., \& Savage, B. 1999, AJ, 117, 2077
\bibitem[]{}Ichikawa, T., van Driel, W., Aoki, T., Soyano, T., Tarusawa, K.,
\& Okamura, S. 1994, ApJ, 433, 645
\bibitem[]{}Iwasawa, K., \& Comastri, A. 1998, MNRAS, 297, 1219
\bibitem[]{}Jablonka, P., Martin, P., \& Arimoto, N. 1996, AJ, 112, 1415
\bibitem[]{}Jacoby, G., Hunter, D., \& Christian, C. 1984, ApJS, 56, 257
\bibitem[]{}Kamaya, H., Mineshige, S., Shibata, K., \& Matsumoto, R. 1996,
ApJ, 458, 25
\bibitem[]{}Kauffmann, G., \& Charlot, S. 1998, MNRAS, 294, 705
\bibitem[]{}Koribalski, B. 1996, in ``Barred Galaxies'', ed. R. Buta, D.
Crocker, \& B. Elmegreen (ASP: San Francisco), p. 172
\bibitem[]{}Kormendy, J. \& Sanders, D. 1992, ApJL, 390, L53
\bibitem[]{}Kraemer, S., Turner, T.J., Crenshaw, D., \& George, I. 1999, ApJ,
519, 69
\bibitem[]{}Kriss, J., Peterson, B., Crenshaw, D., and Zheng, W. 2000,
astro-ph/9912204
\bibitem[]{}Kunth, D., Mas-Hesse, J., Terlevich, E., Terlevich, R.,
Lequeux, J., \& Fall, S.M. 1998, A\&A, 334, 11
\bibitem[]{}Lehnert, M., \& Heckman, T. 1995, ApJS, 97, 89
\bibitem[]{}Lehnert, M. D., \& Heckman, T. M. 1996a, ApJ, 462, 651
\bibitem[]{}Lehnert, M.., \& Heckman, T. 1996b, ApJ, 472, 546
\bibitem[]{}Leitherer, C., \& Heckman, T. M. 1995, ApJS, 96, 9
\bibitem[]{}Leitherer, C., Schaerer, D., Goldader, J., Gonzalez-Delgado, R.,
Robert, C., Kune, D., De Mello, D., Devost, D., \& Heckman, T. 1999, ApJS,
123, 3
\bibitem[]{}Lequeux, J., Kunth, D., Mas-Hesse, J., \& Sargent, W. 1995, A\&A,
301, 18
\bibitem[]{}Lilly, S., Eales, S. Gear, W., Hammer, F., LeFevre, O., Crampton,
D., Bond, J.R., \& Dunne, L. 1999, ApJ, 518, 641
\bibitem[]{}Lo, K.-Y., Cheung, K., Masson, C., Phillips, T., Scott, S. \&
Woody, D. 1987, ApJ, 312, 574
\bibitem[]{}Lutz, D., Veilleux, S., \& Genzel, R. 1999, ApJL, 517, L13
\bibitem[]{}Lynden-Bell, D. 1992, in ``Elements and the Cosmos'', ed.
M. Edmunds \& R. Terlevich (Cambridge University Press: New York), p. 270
\bibitem[]{}Madau, P., Ferguson, H., Dickinson, M., Giavalisco,
M., Steidel, C., \& Fruchter, A. 1996, MNRAS, 283, 1388
\bibitem[]{}Martin, C. 1999, ApJ, 513, 156
\bibitem[]{}Meurer, G., Heckman, T., Leitherer, C., Lowenthal, J.,
\& Lehnert, M. 1997, AJ, 114, 54
\bibitem[]{}Maoz, D. 1995, ApJ, 455, L115
\bibitem[]{}Moran, E., Lehnert, M., \& Helfand, D. 1999, astro-ph/9907036
\bibitem[]{}Morton, D. 1991, ApJS, 77, 119
\bibitem[]{}Mirabel, I.F., \& Sanders, D. 1988, ApJ, 335, 104
\bibitem[]{}Nelson, C., \& Whittle, M. 1996, ApJ, 465, 96
\bibitem[]{}Norman, C., Bowen, D., Heckman, T., Blades, J.C., \& Danly, L.
1996, ApJ, 472, 73
\bibitem[]{}Oliva, E., Origlia, L., Maiolino,R., \& Moorwood, A. 1999,
astro-ph/9908063
\bibitem[]{}Osterbrock, D. 1989, ``Astrophysics of Gaseous Nebulae \&
Active Galactic Nuclei'', (University Science Books: Mill Valley, CA)
\bibitem[]{}Pahre, M., de Carvallo, R., \& Djorgovski, S.G. 1998, AJ, 116, 1606
\bibitem[]{}Perlmutter, S. et al. 1999, ApJ, 517, 565
\bibitem[]{}Pettini, M., Kellogg, M., Steidel, C., Dickinson, M., Adelberger,
K., \& Giavalisco, M. 1998, ApJ, 508, 539
\bibitem[]{}Pettini, M., Steidel, C., Adelberger, K., Dickinson, M., \&
Giavalisco, M. 1999, astro-ph/9908007
\bibitem[]{}Phillips, A. 1993, AJ, 105, 486
\bibitem[]{}Phillips, M. 1976, ApJ, 208, 37
\bibitem[]{}Planesas, P., Mirabel, I.F., \& Sanders, D. 1991, ApJ, 370, 172
\bibitem[]{}Ponman, T., Cannon, D., \& Navarro, J. 1999, Nature, 397, 135
\bibitem[]{}Prada, F., Beckman, J., McKeith, C., Castles, J., \& Greve, A. 1994
ApJL, 423, L35
\bibitem[]{}Read, A., Ponman, T., \& Strickland, D. 1997, MNRAS, 286, 626
\bibitem[]{}Read, A., \& Ponman, T. 1998, MNRAS, 297, 143
\bibitem[]{}Reiss, A. et al. 1998, AJ, 116, 1009
\bibitem[]{}Renzini, A. 1997, ApJ, 488, 35
\bibitem[]{}Renzini, A. 1999, in ``When and How Do Bulges Form \& Evolve?'',
ed. C.M. Carollo, H. Ferguson, \& R. Wyse (Cambridge University Press), in
press (astro-ph/9902108)
\bibitem[]{}Romani, R., \& Maoz, D. 1992, ApJ, 386, 36
\bibitem[]{}Sahu, M., \& Blades, J.C. 1997, ApJL, 484, L125
\bibitem[]{}Sanders, D., \& Mirabel, I.F. 1996, ARA\&A, 34, 749
\bibitem[]{}Sanders, D., Scoville, N., \& Soifer, B.T. 1991, ApJ, 370, 158
\bibitem[]{}Sansom, A., Dotani, T., Okada, K., Yamashita, A., \& Fabbiano, G.
1996, MNRAS, 281, 485
\bibitem[]{}Scarrott, S., Eaton, N., \& Axon, D. 1991, MNRAS, 252, 12
\bibitem[]{}Scarrott, S., Draper, P., Stockdale, D., \& Wolstencroft, R.
1993, MNRAS, 264, 7
\bibitem[]{}Scarrott, S., Draper, P., \& Stockdale, D. 1996, MNRAS, 279, 1325
\bibitem[]{}Savage, B., \& Sembach, K. 1996, ARA\&A, 34, 279
\bibitem[]{}Schmitt, H., Storchi-Bergmann, T., \& Cid-Fernandez, R. 1999,
MNRAS, 303, 173
\bibitem[]{}Sofue, Y., 1998, PASJ, 50, 227
\bibitem[]{}Sofue, Y., Wakamatsu, K.-I., \& Malin, D. 1994, AJ, 108, 2102
\bibitem[]{}Solomon, P., Downes, D., \& Radford S. 1992, ApJ, 387, 55
\bibitem[]{}Spitzer, L. 1968, ``Diffuse Matter in Space'', (Interscience:
New York)
\bibitem[]{}Steidel, C., Adelberger, K., Giavalisco, M., Dickinson, M., \&
Pettini, M. 1999, ApJ, 519, 1
\bibitem[]{}Stickel, M., Lemke, D., Haas, M., Mattila, K., Haikala, H. 1998,
A\&A, 329, 55
\bibitem[]{}Strel'nitskii and Sunyaev 1973
\bibitem[]{}Strickland, D. K. 1998, Ph.D. Dissertation, The University of
Birmingham
\bibitem[]{}Strickland, D. K., Ponman, T. J., \& Stevens, I. R. 1997, A\&A,
320, 378
\bibitem[]{}Suchkov, A. A., Balsara, D. S., Heckman, T. M., \& Leitherer, C.
1994, ApJ, 430, 511
\bibitem[]{}Suchkov, A., Berman, V., Heckman, T., \& Balsara, D.
1996, ApJ, 463, 528
\bibitem[]{}Tenorio-Tagle, G., \& Munoz-Tunon, C. 1998, MNRAS, 293, 299
\bibitem[]{}Tomisaka, K., \& Bregman, J. 1993, PASJ, 45, 513
\bibitem[]{}Trager, S., Worthey, G., Faber, S., Burstein, D., \&
Gonzalez, J. 1998, ApJS, 116, 1
\bibitem[]{}Vaceli, M., Viegas, S., Gruenwald, R., \& De Souza, R. 1997,
AJ, 114, 1345
\bibitem[]{}Veilleux, S., Kim, D.-C., Sanders, D., Mazzarella, J.,
\& Soifer, B.T. 1995, ApJS, 98, 171
\bibitem[]{} Weaver, R., McCray, R., Castor, J., Shapiro, P., \& Moore, R.
1977, ApJ, 218, 377
\bibitem[]{}Weedman, D. 1983, ApJ, 266, 479
\bibitem[]{}Whittle, M. 1992, ApJ, 387, 109
\bibitem[]{}Wyse, R., \& Silk, J. 1985, ApJL, 296, L1
\bibitem[]{}Young, J., et al. 1995, ApJS, 98, 219
\bibitem[]{}Zaritsky, D. 1994, AJ, 108, 1619
\bibitem[]{}Zezas, A., Georgantopoulos, I., \& Ward, M. 1998, MNRAS, 301, 915

\end{thebibliography}
